\documentclass[prl,aps,twocolumn,superscriptaddress,notitlepage, nobalancelastpage]{revtex4-1}

\usepackage[utf8]{inputenc}
\usepackage[T1]{fontenc} 
\usepackage{lmodern} 
\usepackage{microtype} 
\usepackage{graphicx}
\usepackage{dcolumn}
\usepackage{enumerate,amsthm,amssymb,color}
\usepackage{amsfonts}
\usepackage{amsmath}
\usepackage{mathtools}
\usepackage{dsfont}

\usepackage{subcaption}
\usepackage{ragged2e}
\DeclareCaptionJustification{justified}{\justifying}
\usepackage{bbm}
\usepackage{bm}
\usepackage{synttree}
\usepackage{float}
\usepackage{bbold}
\usepackage{braket}
\usepackage{xcolor}
\usepackage{gensymb}

\usepackage{booktabs}
\usepackage{upgreek, eurosym}
\usepackage[colorlinks=true]{hyperref}

\newcommand{\abs}[1]{\left|#1\right|}

\newcommand{\ketbra}[2]{|#1\rangle\!\langle#2|}

\DeclareMathOperator{\Tr}{Tr}

\newcommand{\us}{\mskip3mu\upmu\textrm{s}}

\newcommand{\money}{\text{\euro}}

\graphicspath{{images/}}

\begin{document}

\title{Semi-device-independent quantum money with coherent states}

\author{Mathieu Bozzio}\affiliation{LIP6, CNRS, Sorbonne Universit\'e, 75005 Paris, France}\affiliation{LTCI, T\'el\'ecom ParisTech, Universit\'e Paris-Saclay, 75013 Paris, France}
\author{Eleni Diamanti}\affiliation{LIP6, CNRS, Sorbonne Universit\'e, 75005 Paris, France}
\author{Fr\'ed\'eric Grosshans}\affiliation{Laboratoire Aim\'e Cotton, CNRS, Universit\'e Paris-Sud, ENS Cachan, Universit\'e Paris-Saclay, 91405 Orsay Cedex, France}
\affiliation{LIP6, CNRS, Sorbonne Universit\'e, 75005 Paris, France}

\begin{abstract}
The no-cloning property of quantum mechanics allows unforgeability of quantum banknotes and credit cards. Quantum credit card protocols involve a bank, a client and a payment terminal, and their practical implementation typically relies on encoding information on weak coherent states of light. Here, we provide a security proof in this practical setting for semi-device-independent quantum money with classical verification, involving an honest bank, a dishonest client and a potentially untrusted terminal. Our analysis uses semidefinite programming in the coherent state framework and aims at simultaneously optimizing over the noise and losses introduced by a dishonest party. We discuss secure regimes of operation in both fixed and randomized phase settings, taking into account experimental imperfections. Finally, we study the evolution of protocol security in the presence of a decohering optical quantum memory and identify secure credit card lifetimes for a specific configuration.
\end{abstract}

\date{\today}

\maketitle

In contrast to classical physics, quantum mechanical systems have a no-cloning property \cite{WZ:nature82}: it is impossible to make a perfect copy of a quantum object in an unknown state. This property was used by Wiesner in his proposal to mint unforgeable quantum money \cite{Wie:acm83}, giving birth to the field of quantum cryptography \cite{BB84,GRT:RMP02,SBP:RMP09}.
%
%
The original idea involved a bank encoding a secret classical key into a sequence of two-level quantum states (qubits) stored in a quantum memory and handed to a client.
The secret key specifies the basis in which each qubit is encoded, ensuring that a forger ignoring the basis in which to measure it will destroy information. This will then trigger incorrect measurement outcomes when the bank will verify the validity of the banknote. Such a scheme may be impractical over long distances due to a potentially lossy and noisy transmission of the quantum states between the client and the bank. It was also shown to be vulnerable to adaptive attacks, where a counterfeiter can use the same banknote several times \cite{BNS:QIC16}. An alternative protocol with verification using classical communication was first proposed in \cite{Gav:ccc12} and extended to practical, noise-tolerant schemes in \cite{PY+:pnas12,GK:tqc15,AA:pra17}.

Although quantum key distribution protocols have been widely studied and implemented \cite{DL+:npjQI16}, quantum money has not yet seen the same experimental progress, essentially because of the difficulty in implementing efficient quantum storage devices 
\cite{HE+:jmo16}. However, the experimental interest in quantum money has grown recently, with demonstration of forgery of quantum banknotes \cite{BC+:npjQI17} and implementation of weak coherent state- based quantum credit card schemes, secure in a trusted terminal scenario \cite{BOV:npj18,GAA:pra18}, in the prospect of near-future implementations with a quantum memory. These require new security proofs tackling the optimal cloning of coherent states, differing from qubit-based quantum money and also quantum key distribution proofs.

In quantum cryptography, semi-device-independent frameworks have been developed in order to limit the needed assumptions to ensure security. While not as stringent as full device independence \cite{ABG:PRL07}, this approach allows for practical security and performance while making fewer assumptions on the implementation than usual security proofs. This includes assumptions on the detectors \cite{ML:QIC12,BCW:PRA12,LCQ:PRL12,BP:PRL12}, the dimensionality of the quantum states \cite{LVB:PRA11,PB:PRA11,CM:AIP13} and other parameters \cite{VHWC:Qua17}. For quantum money, semi-device-independence relates to scenarios where one does not trust the terminal, as in this work and \cite{HS:arx18,JBC:arx18}, along with scenarios where the state preparation \cite{HS:arx18} or the terminal is trusted but imperfectly characterized.

In this work, we derive a quantum money security proof which incorporates semi-device-independence to deal with both trusted and untrusted payment terminals in the presence of experimental imperfections. We do so by extending the semidefinite programming (SDP) techniques from \cite{MVW:tqc12,W:LN11,VB:SIAM96} to the coherent state framework and using the squashing model from \cite{BML:prl08,GBN:pra14}. We also adapt our proof to study the effect of a decohering quantum memory. We remark that recent and concurrent work by Horodecki and Stankiewicz \cite{HS:arx18} also studies semi-device-independent quantum money, in a stronger threat model than here (scenario \textit{(iv)} of Table \ref{tab:DH}), but without our focus on realistic implementations.

\section{I. Protocol and correctness}

We consider the qubit scheme introduced by Wiesner \cite{Wie:acm83} in the classical verification setting of \cite{Gav:ccc12,PY+:pnas12,GK:tqc15,AA:pra17}.
In this three-party quantum money scheme, the mint generates a random secret classical key $k^{(s)}$ and encodes it according to a secret classical basis key $b^{(s)}$. The quantum credit card state associated to public serial number $s$ is then written as $\ket{\money^{(k,b)}}=\bigotimes_{j=1}^{n}\ket{\psi_{j}^{(k,b)}}$, where $\ket{\psi_{j}^{(k,b)}}\in\{\ket{+},\ket{+i},\ket{-},\ket{-i}\}$. More specifically, bit $k^{(s)}_j$ is encoded in the $\sigma_x$ basis when $b^{(s)}_j=0$, and in the $\sigma_y$ basis when $b^{(s)}_j=1$.

The mint stores $\ket{\money^{(k,b)}}$ in a quantum memory and hands it to a client. When a transaction must be performed, the merchant's honest terminal measures each of the $n$ qubits of $\ket{\money^{(k,b)}}$ in a basis dictated by a challenge question randomly chosen by the bank. For a single qubit state, the challenge $c_{0}$ (resp. $c_{1}$) reads: \textit{Give the correct measurement outcome if the qubit is encoded in the $\sigma_{x}$ (resp. $\sigma_{y}$) basis, and provide any outcome if the qubit is encoded in the $\sigma_{y}$ (resp. $\sigma_{x}$) basis.} The terminal measures the qubit in the basis associated with the given challenge, which provides the honest success probability or \textit{correctness} $c=1$. The answers corresponding to the measurement results are sent in the form of a classical bit string to the bank, which compares it with $k^{(s)}$ and accepts the credit card only if all the measurement outcomes coincide with $k^{(s)}$.

We now consider the same honest protocol in which qubit states are mapped onto two-mode weak coherent states as:
\begin{equation}
\begin{aligned}
\ket{0} &\rightarrow \ket{\alpha}\otimes\ket{\text{vac}}
           & \ket{1} &\rightarrow \ket{\text{vac}}\otimes\ket{-\alpha}\\
\ket{+} &\rightarrow \ket{\tfrac{\alpha}{\sqrt{2}}}\otimes\ket{\tfrac{\alpha}{\sqrt{2}}}
           &\ket{-} &\rightarrow \ket{\tfrac{\alpha}{\sqrt{2}}}\otimes\ket{-\tfrac{\alpha}{\sqrt{2}}}\\
\ket{+i}&\rightarrow \ket{\tfrac{\alpha}{\sqrt{2}}}\otimes\ket{i\tfrac{\alpha}{\sqrt{2}}}
           &\ket{-i}&\rightarrow \ket{\tfrac{\alpha}{\sqrt{2}}}\otimes\ket{-i\tfrac{\alpha}{\sqrt{2}}}, 
\end{aligned}
\label{pop}
\end{equation}
where $\alpha$ is the coherent state amplitude and $\ket{\text{vac}}$ denotes the vacuum state. Such a mapping is typically used for polarization or time-bin encoding, 
where the $\ket{0}$ component is mapped onto the first mode and the $\ket{1}$ component is mapped onto the second \cite{LP:QIC07}.
When dealing with polarization, the honest terminal measures each of the $n$ credit card states in the basis which answers either $c_{0}$ or $c_{1}$ by typically rotating a half or quarter waveplate.
It then outputs $(1-f_{h}) n$ measurement outcomes, where $f_{h}\approx e^{-\eta_d \mu}$ represents the honest losses assuming a weak coherent light source with average photon number per pulse $\mu=|\alpha|^2$, unit channel transmission efficiency, and threshold single-photon detectors with quantum efficiency $\eta_d$. When no detection occurs, the terminal reports a flag, denoted by $\varnothing$.
For large sample sizes, the $n$-state challenge is then satisfied only if the total number of no-detection reports is equal to $f_h n$. The multi-photon component of coherent states may also trigger clicks on both detectors at the same time. An adversary may actually exploit this property to boost his cheating probability. Following the methods used in \cite{BML:prl08}, clicks on both detectors are randomly mapped to a single click as either a $0$ or $1$. This allows to use a squashing model to securely map the infinite-dimensional threshold detection POVMs to a finite dimensional Hilbert space.

\begin{table}
\begin{ruledtabular}
\begin{tabular}{cccccc}
 & Mint & Client & Terminal
 & Bank  & Parameter \\
\hline
\textit{(i)} & H & H & H & H & correctness c\\
\textit{(ii)} & H & D & H & H & error rate $e$\\
\textit{(iii)} & H & D/H & D & H & error rate $e$\\
\textit{(iv)} & D & D/H &D & H &  N/A\\
\textit{(v)} & D & D/H &D/H & D &  N/A
\end{tabular}
\end{ruledtabular}
\caption{\label{tab:DH}Scenarios for quantum money with classical verification in terms of honest (H) and dishonest (D) parties. Cases denoted by D/H are indistinguishable to the bank. In this work, \textit{(ii)} is studied in a semi-device-independent regime, since the squashing model allows to strongly limit the assumptions on the terminal detectors, while both \textit{(iii)} and \textit{(iv)} are by definition semi-device-independent.
Here, we do not study \textit{(iv)}, recently defined and analyzed in \cite{HS:arx18}, or the unrealistic scenario \textit{(v)}.}
\end{table}

\section{II. Security}

\subsection{A. Principle and proof outline}

Table \ref{tab:DH} shows the possible security scenarios for quantum money schemes. A successful forging attack consists in answering two challenges correctly at the same time, corresponding to extracting twice the original amount of money in one's possession. As the last four states from Eq.~($\ref{pop}$) are identical on the first mode, we may reduce our security analysis to the single state $\ket{\alpha_k}=\ket{i^k\frac{\alpha}{\sqrt{2}}}$ with $k\leqslant 3 $,
before extending it to $n$ states.
In scenario \textit{(ii)}, an attack is materialized by the creation of two copies of the quantum credit card state, both being accepted by the bank when measured by two separate trusted terminals.
In 
scenario \textit{(iii)}, an attack is materialized by the communication of two classical strings by two untrusted terminals to the bank, which accepts both of them.
In a coherent state implementation, the adversary may modify one or both of the following parameters: losses $f_{d}$ (probability of a projection onto the vacuum state), and error rate $e$. The bank may detect an attack when $f_{d}>f_{h}$ or when the measured error rate $e$ upon verification is larger than expected. Given average photon number $\mu$, we use SDP techniques \cite{W:LN11,VB:SIAM96} to first minimize the losses that the adversary must introduce in \textit{(ii)} or declare in \textit{(iii)} to succeed with probability $(1-e)$.
We can then identify 
the range of $\mu$ for which $f_{d}>f_{h}$.  We will use Choi's theorem (see Appendix A.1 for details) to optimize over the best adversarial linear cloning map.
For \textit{(ii)}, the figure of merit for the optimization is based on the measurements of the two trusted terminals. For \textit{(iii)}, the figure of merit becomes the acceptance of classical data by the bank. We then show how this single state analysis gives a bound for the $n$-state proof. We also note the following useful property from \cite{MVW:tqc12}, proven in Appendix A.2: given $\ket{\psi_1}\in\mathcal{H}_1^d$, $
\ket{\psi_3}\in\mathcal{H}_3^{d'}$, and Choi--Jamio\l kowski operator $J(\Lambda)$ associated to map $\Lambda$, we have
\begin{equation}
    \bra{\psi_3}\Lambda(\ket{\psi_1}\bra{\psi_1})\ket{\psi_3}
    =\bra{\psi_3}\otimes\bra{\overline{\psi_1}} J(\Lambda)  \ket{\psi_3}\otimes\ket{\overline{\psi_1}},
    \label{luke}
\end{equation}
where the overline denotes complex conjugation.

\subsection{B. Trusted terminal}

\begin{figure*}
\begin{subfigure}{0.5\textwidth}
  {\centering
  \includegraphics[width=1\linewidth]{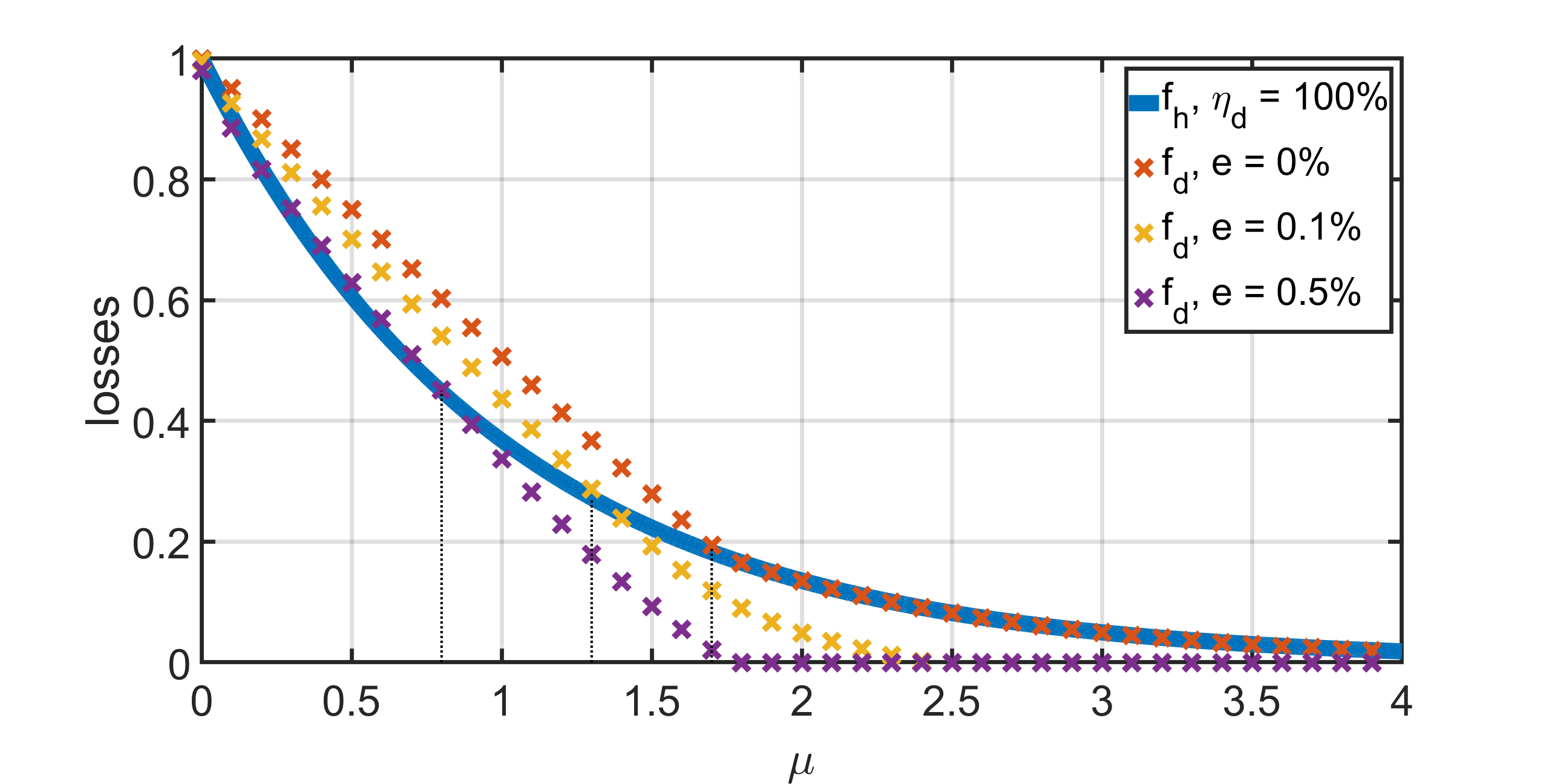}
  \caption{Trusted, fixed phase, $\eta_d=100\%$}
  \label{chuck1}}
\end{subfigure}%
\begin{subfigure}{0.5\textwidth}
  {\centering
  \includegraphics[width=1\linewidth]{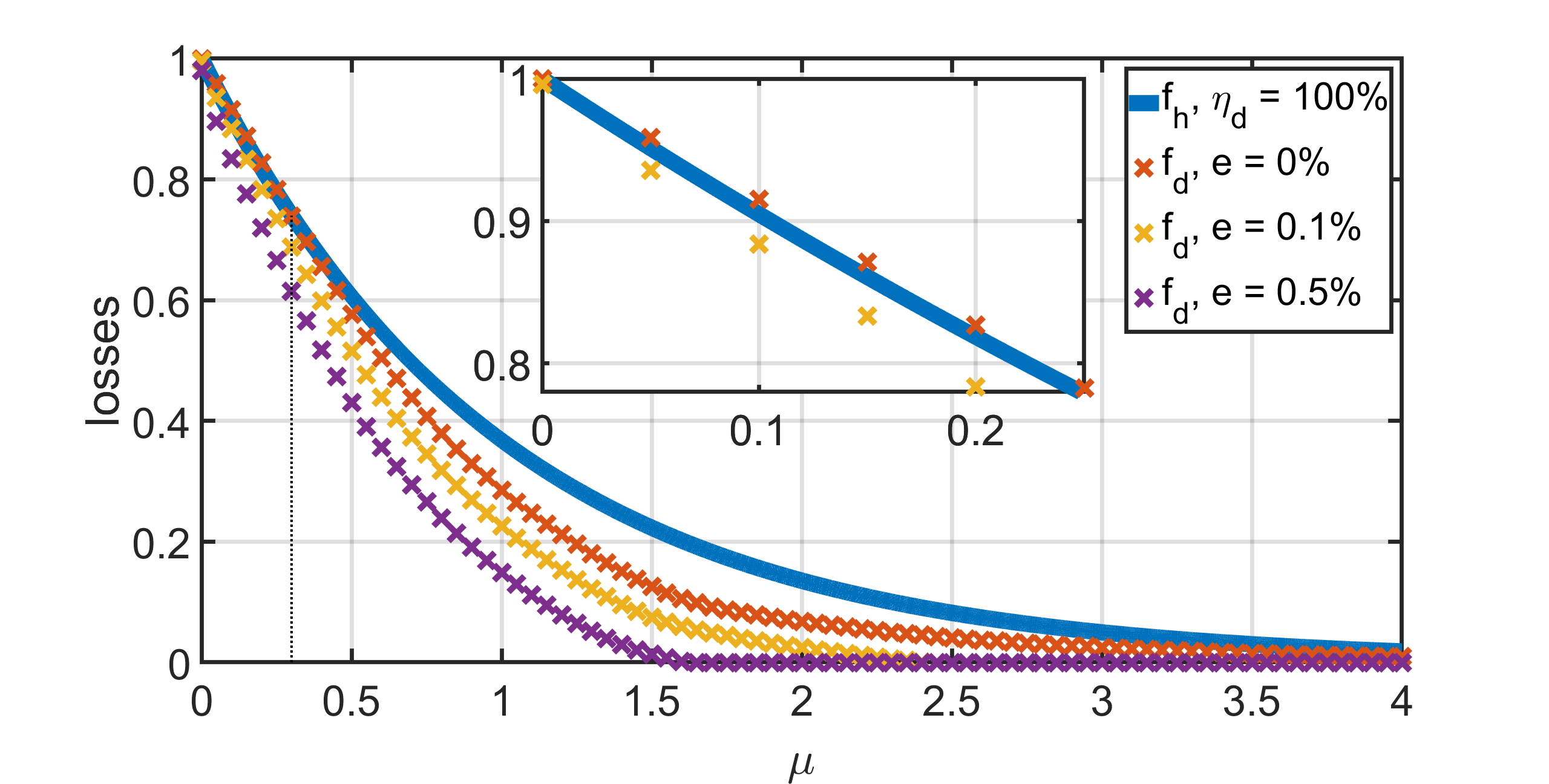}
  \caption{Untrusted, fixed phase, $\eta_d=100\%$}
  \label{chuck2}}
\end{subfigure}
\begin{subfigure}{0.5\textwidth}
  {\centering
  \includegraphics[width=1\linewidth]{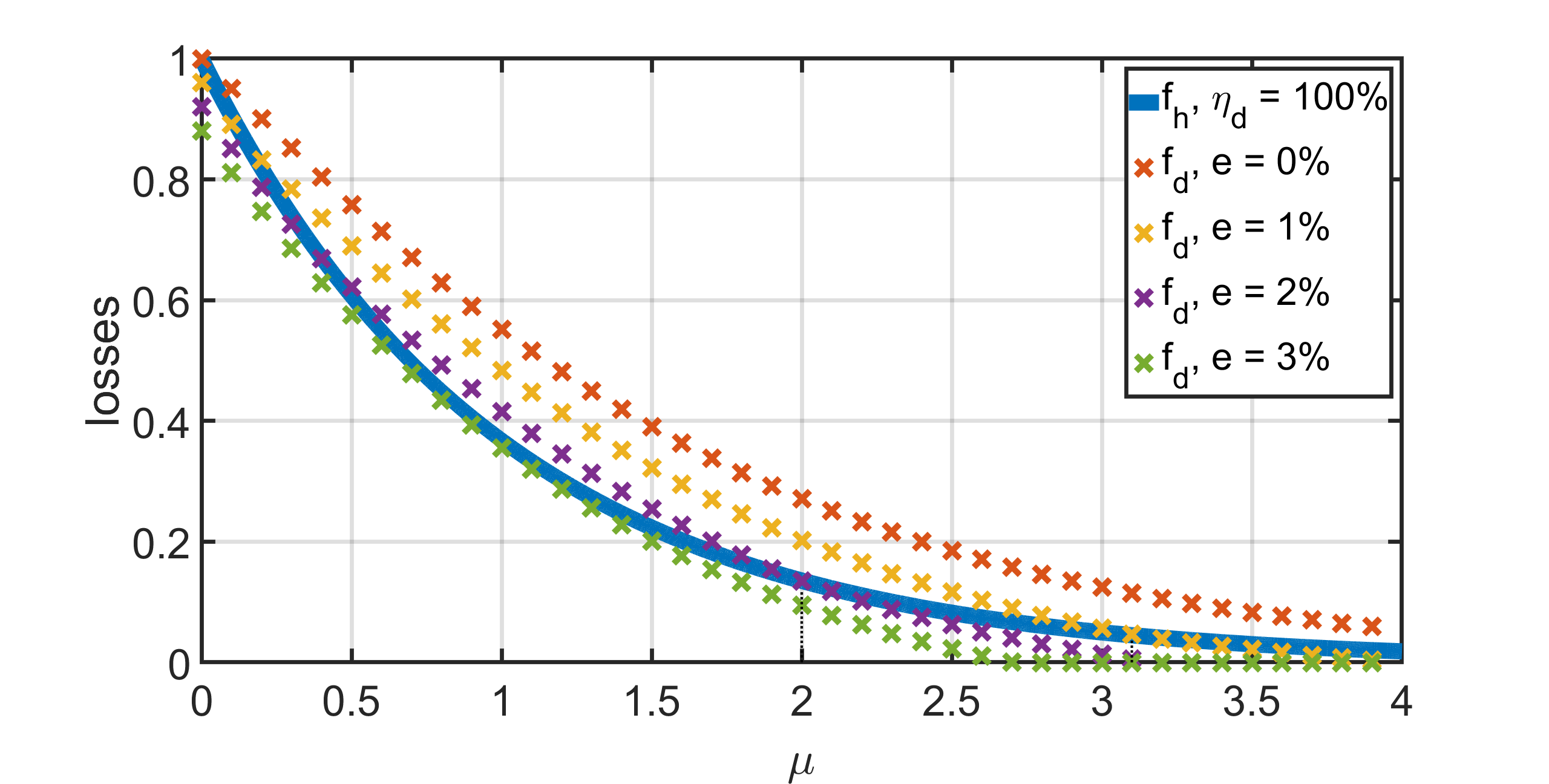}
  \caption{Trusted, randomized phase, $\eta_d=100\%$}
  \label{chuck3}}
\end{subfigure}%
\begin{subfigure}{0.5\textwidth}
  {\centering
  \includegraphics[width=1\linewidth]{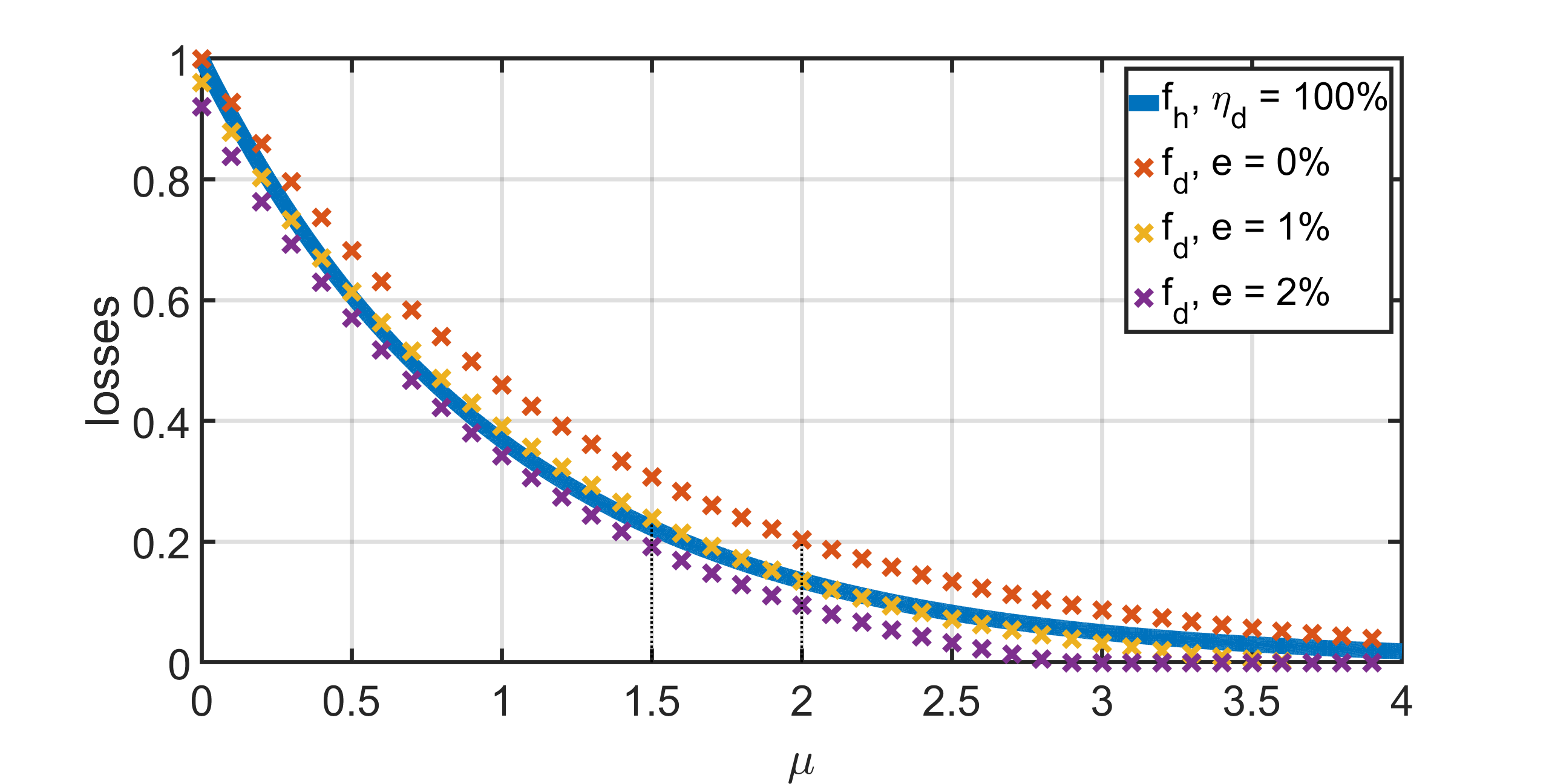}
  \caption{Untrusted, randomized phase, $\eta_d=100\%$}
  \label{chuck4}}
\end{subfigure}
\begin{subfigure}{0.5\textwidth}
  {\centering
  \includegraphics[width=1\linewidth]{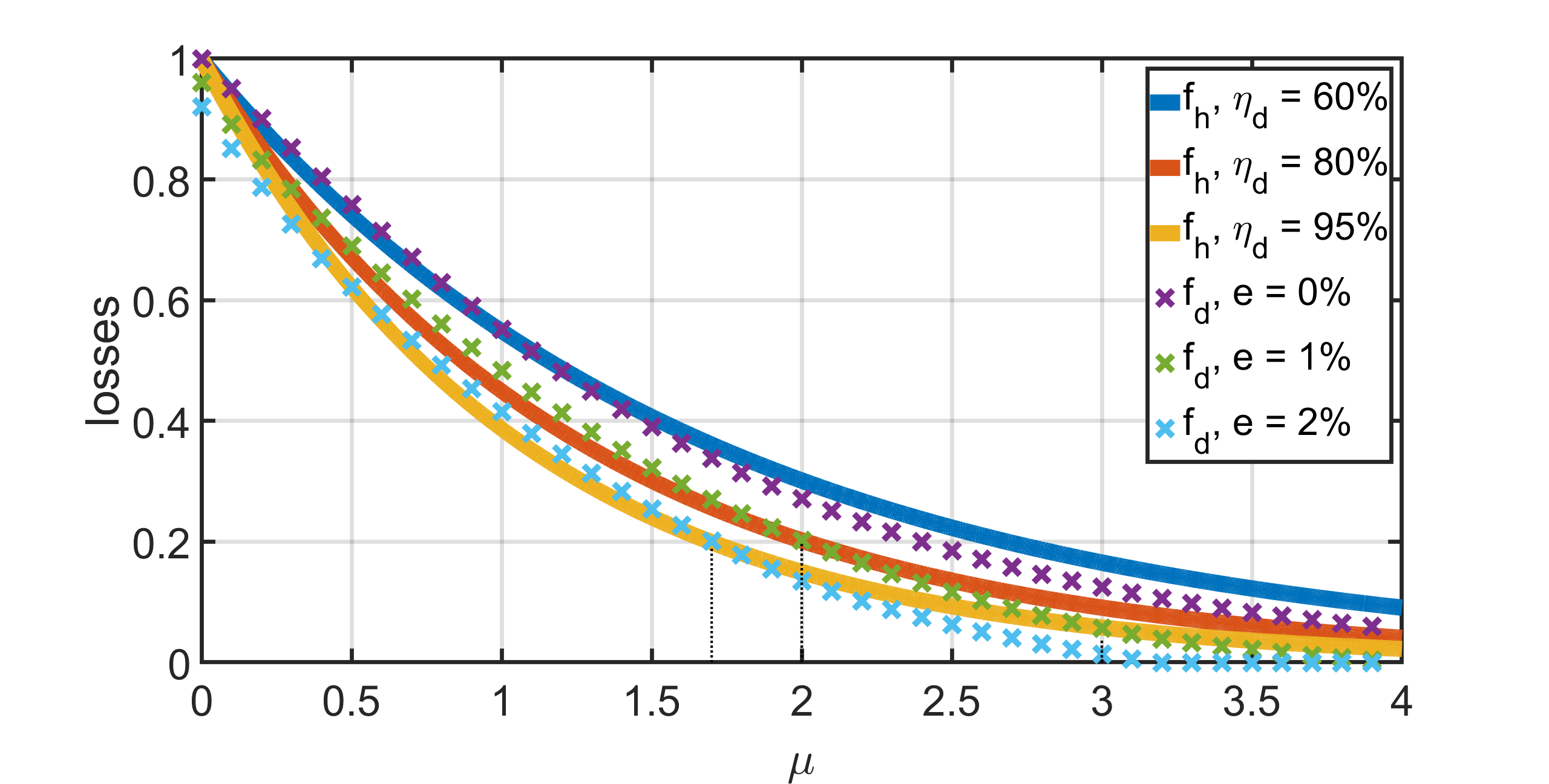}
  \caption{Trusted, randomized phase, $\eta_d<100\%$}
  \label{chuck5}}
\end{subfigure}%
\begin{subfigure}{0.5\textwidth}
  {\centering
  \includegraphics[width=1\linewidth]{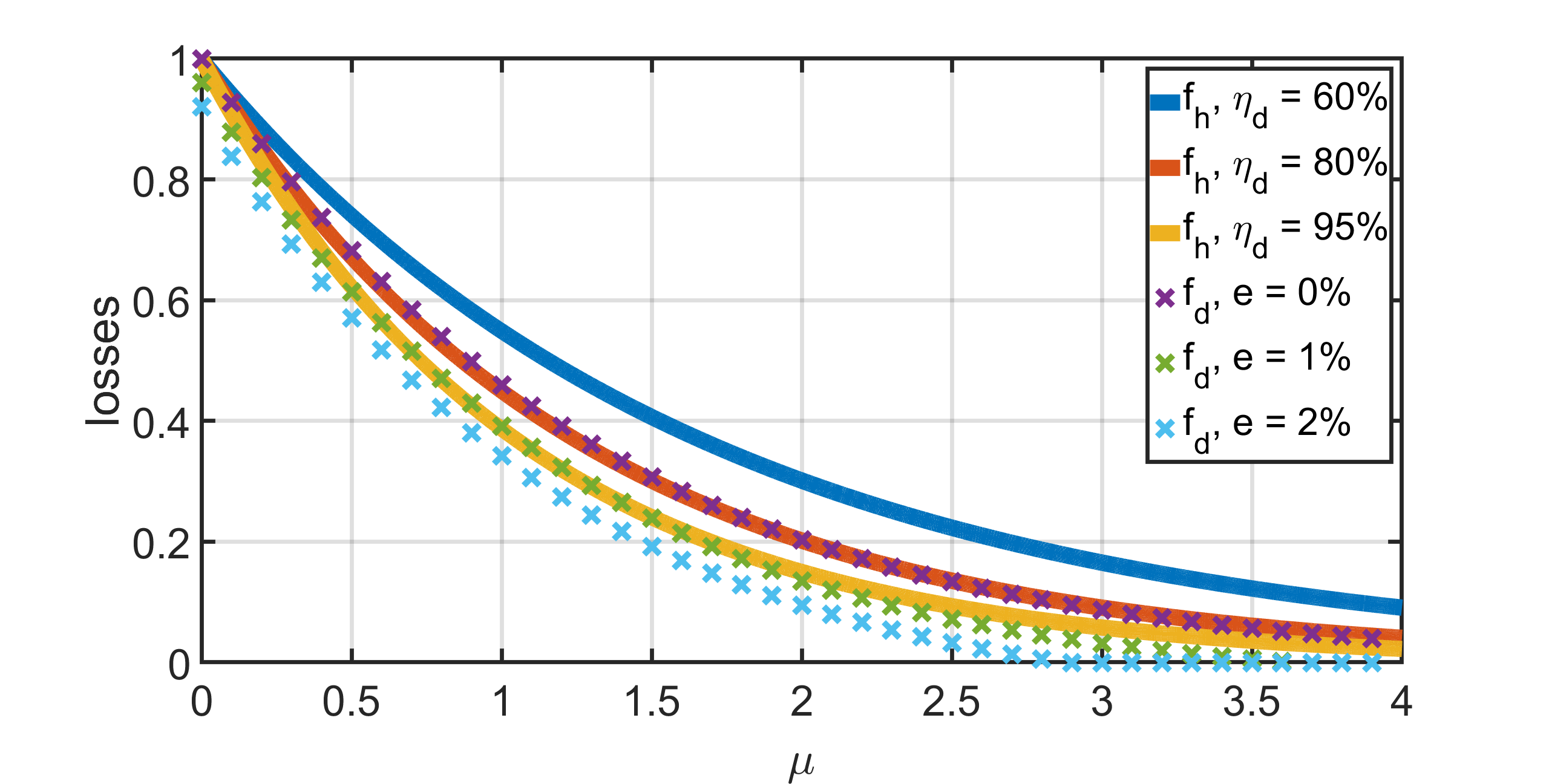}
  \caption{Untrusted, randomized phase, $\eta_d<100\%$}
  \label{chuck6}}
\end{subfigure}
\caption{Optimal numerical solutions of problem (\ref{bobby}) as a function of average photon number $\mu$ for different values of error rate $e$ and detection efficiency $\eta_d$, for both trusted and untrusted terminals. Solid lines correspond to the honest losses $f_{h}=e^{-\eta_d\mu}$. Points indicate the losses $f_{d}$ that a dishonest party must induce in order to succeed with error $e$. Graphs are plotted from top to bottom following the order of the legend. The protocol is secure in regions where $f_{d}>f_{h}$. We used the \textit{SDPT3} solver \cite{SDPT3:a, SDPT3:b} of the \textit{CVX} \cite{cvx, gb08} software.}
\label{chuck}
\end{figure*}

We shall first study the trusted terminal scenario \textit{(ii)}. In the single qubit case, the minimum adversarial error probability is the same as in Wiesner's original quantum verification scheme, namely $e=1/4$ \cite{Wie:acm83,MVW:tqc12}. When dealing with the coherent states from Eq.~(\ref{pop}), we use the existence of a squashing model for our threshold detector measurement setup, originally proven for coherent implementations of BB84 \cite{BML:prl08}. By imposing a condition on the terminal's postprocessing, consisting of assigning a random measurement outcome to any double click, this model allows to express the infinite-dimensional measurement operators in a 3-dimensional space spanned by $\{\ket{0},\ket{1},\ket{\varnothing}\}$, which greatly simplifies the security analysis. Let $\Lambda$ be the optimal adversarial map which produces two copies (living in $\mathcal{H}_1\otimes\mathcal{H}_2$) of the original quantum credit card state $\rho_{\text{mint}} = \frac{1}{4}\sum_{k=0}^3\ket{\alpha_k}\bra{\alpha_k}$ (living in  $\mathcal{H}_{\text{mint}}$). The state $\rho_{\text{mint}}$ may be expressed in a $4$-dimensional orthonormal basis corresponding to $\text{span}\{\ket{\alpha_k}\}$, as shown in Appendix B.1.  The probability that a trusted terminal declares an incorrect measurement outcome for credit card $1$ (resp.\@ $2$) is given by the trace of $\sum_{k=0}^3\left(\frac{1}{2}\ket{\beta_{k}^\perp}\bra{\beta_{k}^\perp}\otimes \mathbb{1}\right) \Lambda(\frac{1}{4}\ket{\alpha_k}\bra{\alpha_k})$,
(resp.\@ of $\sum_{k=0}^3\left(\mathbb{1}\otimes\frac{1}{2}\ket{\beta_{k}^\perp}\bra{\beta_{k}^\perp}\right) \Lambda(\frac{1}{4}\ket{\alpha_k}\bra{\alpha_k}) $,
where $\ket{\beta_{k}}$ is the squashed qubit associated with the original state $\ket{\alpha_k}$, i.e.\@ $\ket{\beta_0} = \ket{+}$, $\ket{\beta_1} = \ket{+i}$, $\ket{\beta_2} = \ket{-}$, $\ket{\beta_3} = \ket{-i}$, and $\ket{\beta_{k}^\perp}$ is its orthogonal qubit state. The factor $1/4$ indicates that each $\ket{\alpha_k}$ is equally likely to occur, while $1/2$ accounts for the trusted terminal's random measurement basis choice. Using Eq.~($\ref{luke}$), we may then rewrite these expressions as $\Tr\left(E_1(\mu)J(\Lambda)\right)$ and $\Tr\left(E_2(\mu)J(\Lambda)\right)$, where $E_1(\mu)$ and $E_2(\mu)$ are the \textit{error operators},
\begin{equation}
\begin{aligned}
   E_1(\mu)=&\frac{1}{4}\sum_{k=0}^{3}\frac{1}{2} \ket{\beta_{k}^\perp}\bra{\beta_{k}^\perp}\otimes\mathbb{1}\otimes \ket{\overline{\alpha_k}}\bra{\overline{\alpha_k}}\\
   E_2(\mu)=&\frac{1}{4}\sum_{k=0}^{3}\mathbb{1}\otimes\frac{1}{2} \ket{\beta_{k}^\perp}\bra{\beta_{k}^\perp}\otimes \ket{\overline{\alpha_k}}\bra{\overline{\alpha_k}},
\end{aligned}
\end{equation}
and $\mu=|\alpha|^2$ is the average photon number in a pulse.
Following a similar method, the probability that terminal 1 (resp. 2) registers a no-detection event on credit card $1$ (resp. $2$) reads $\Tr\left(L_1(\mu)J(\Lambda)\right)$ (resp. $\Tr\left(L_2(\mu)J(\Lambda)\right)$), where $L_1(\mu)$ and $L_2(\mu)$ are the \textit{loss operators}, which contain the projection onto the state $\ket{\varnothing}$:
\begin{equation}
\begin{aligned}
L_1(\mu) = \frac{1}{4}\sum_{k=0}^3 \ket{\varnothing}\bra{\varnothing}\otimes\mathbb{1} \otimes\ket{\overline{\alpha_k}}\bra{\overline{\alpha_k}} \\
L_2(\mu) = \frac{1}{4}\sum_{k=0}^3
\mathbb{1}\otimes\ket{\varnothing}\bra{\varnothing} \otimes\ket{\overline{\alpha_k}}\bra{\overline{\alpha_k}}.
\label{will}
\end{aligned}
\end{equation}
We now search for the optimal cloning map $\Lambda$ that minimizes the losses that the adversary must introduce on both credit cards for a given error rate $e$. We cast this problem in the following SDP for a card with a single state,
\begin{equation}
\begin{aligned}
\min &&& \Tr\left(L_1(\mu) J(\Lambda) \right)\\
\text{s.t.}  &&& \Tr_{\mathcal{H}_1\otimes\mathcal{H}_2}\left(J(\Lambda)\right) = \mathbb{1}_{\mathcal{H}_{\text{mint}}} \\
&&&\Tr\left(E_1(\mu) J(\Lambda) \right) = e \\
&&& \Tr\left(E_1(\mu) J(\Lambda) \right) \geqslant \Tr\left(E_2(\mu) J(\Lambda) \right) \\
&&& \Tr\left(L_1(\mu) J(\Lambda) \right) \geqslant \Tr\left(L_2(\mu) J(\Lambda) \right) \\
&&& J(\Lambda) \geqslant 0.
\end{aligned}\label{bobby}
\end{equation}
The first constraint imposes that $\Lambda$ is trace-preserving, the second imposes error $e$ when card $1$ is measured by terminal $1$, the third and fourth impose that the error and losses on card $1$ are at least equal to those on card $2$, and the fifth imposes that $\Lambda$ is completely positive. Solving (\ref{bobby}) numerically provides the results in Fig. \ref{chuck1}: it is impossible for an adversary to succeed with zero error ($e=0\%$) without introducing any excess losses ($f_d>f_h$) when $\mu<1.7$. The protocol may therefore be implemented securely in this range of $\mu$, since the excess losses will allow the bank to detect an attack. Secure regions of operation for other values of error $e$ are also displayed in Fig. \ref{chuck1}.

In Appendix C, we extend problem (\ref{bobby}) to $n$ states and provide numerical evidence that the optimal solution does not change in this case, namely the adversary cannot decrease $f_d$ by correlating the $n$ states.

\subsection{C. Untrusted terminal}

In the untrusted terminal scenario \textit{(iii)}, the adversary aims to provide two classical outcome strings from two different untrusted terminals which are both accepted by the bank. The minimum error in the qubit case yields $e = 1/8$ \cite{MVW:tqc12} (attained with the strategy provided in Appendix D). In the coherent state framework, we recast ($\ref{bobby}$) with newly defined error and loss operators:
\begin{equation}
  \begin{aligned}
    E_1(\mu) = \frac{1}{16}\sum_{i,j=0}^{1}&\smashoperator[r]{\sum_{k\in\{i,i+2\}}} \ket{a_{ik}^{\perp}}\bra{a_{ik}^{\perp}}\otimes \mathbb{1}\\
    &\hspace{1cm}  \otimes\ket{c_i,c_j,\overline{\alpha_k}}\bra{c_i,c_j,\overline{\alpha_k}}\\
    L_1(\mu) = \frac{1}{16}\sum_{i,j=0}^{1}\sum_{k=0}^3& \ket{\varnothing}\bra{\varnothing}\otimes\mathbb{1}\\ &\hspace{1cm}\otimes\ket{c_i,c_j,\overline{\alpha_k}}\bra{c_i,c_j,\overline{\alpha_k}},
    \label{fiddl}
\end{aligned}
\end{equation}
and similarly for $E_2(\mu)$ and $L_2(\mu)$. We use \textit{braket} notation to denote the correct classical answer $\ket{a_{ik}}$ to challenge $\ket{c_{i}}$, given state $\ket{\alpha_k}$. These vectors are all orthogonal to one another, and live in a 3-dimensional space spanned by classical answers $\{\ket{a_0},\ket{a_1},\ket{\varnothing}\}$, where the last vector corresponds to a classical no-detection flag. We label the orthogonal (wrong) answer as $\ket{a_{ik}^\perp}$. Figure \ref{chuck2} displays the optimal solutions as a function of $\mu$: an errorless protocol is impossible without increasing the fraction of declared no-detection flags with respect to the honest fraction $f_h$, although this increase is extremely small compared to the trusted terminal setting (see figure inset).

\section{III. Parameter analysis}

The small adversarial losses and tight noise tolerance observed in Fig. \ref{chuck2} may be increased by replacing the pure states $\{\ket{\alpha_k}\}$ (given in Appendix B.1) with phase-randomized states $\rho_k$ (expressions given in Appendix B.2). Phase randomization is commonly used in quantum key distribution implementations to increase the security and obtain higher key rates \cite{LP:calt05,ZQL:apl07,YLD:PRAPP14}. Numerical solutions to (\ref{bobby}) for such states are displayed in Figs. \ref{chuck3} and \ref{chuck4} for trusted and untrusted terminals respectively. We observe that the range of $\mu$ for which security can be shown in practice is considerably extended in this case.

It is also interesting to analyze our results in this phase-randomized setting for finite detection efficiency $\eta_d$. Figures \ref{chuck5} and \ref{chuck6} show that security may be achieved in the trusted terminal scenario using state-of-the-art single-photon detectors \cite{H:NP09,LMN:OE08}, depending also on the target error rate, while the untrusted scenario puts much more stringent constraints on the required devices.

We also remark that in Appendix E, we provide an alternative SDP to (\ref{bobby}) which allows to derive $e$ given a fixed $\mu$ and detection efficiency $\eta_d$.

\section{IV. Decohering quantum memory}

\begin{figure}
\begin{subfigure}{0.5\textwidth}
   \centering
   \includegraphics[width=1\linewidth]{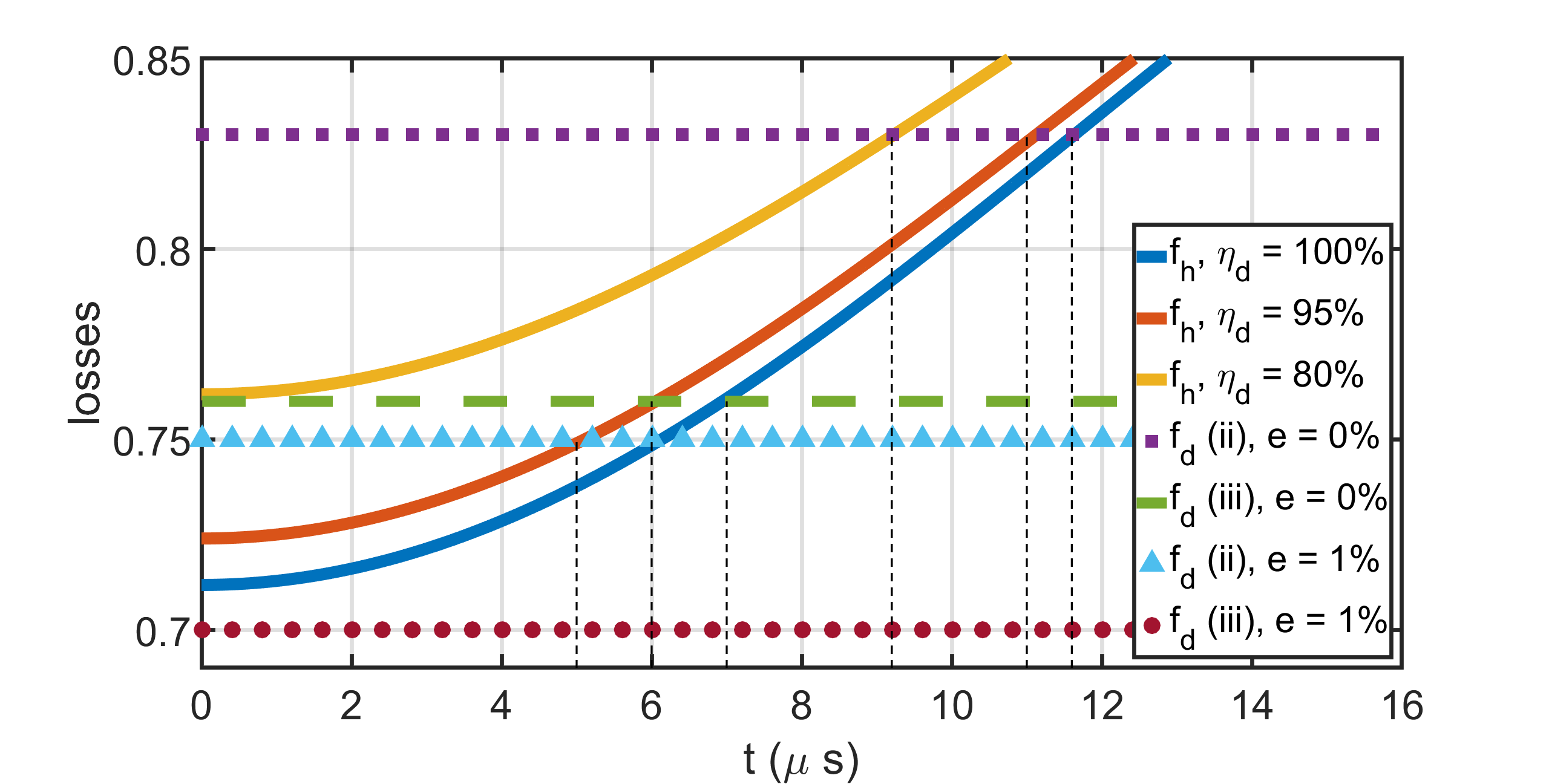}
   \caption{$\mu=0.50$}
   \label{fig:mem1}
 \end{subfigure}
 \begin{subfigure}{0.5\textwidth}
   \centering
   \includegraphics[width=1\linewidth]{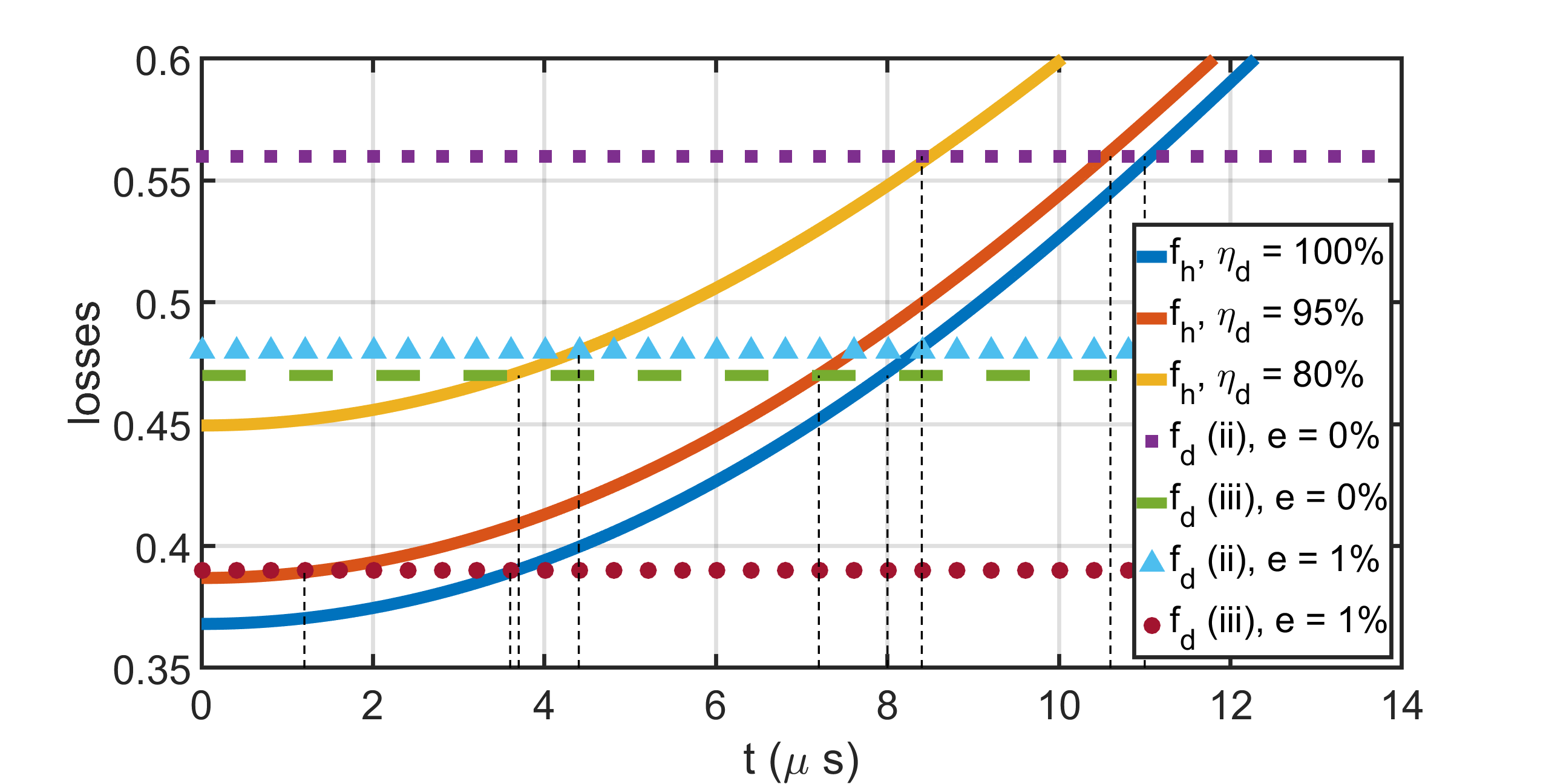}
   \caption{$\mu=1.50$}
   \label{fig:mem2}
 \end{subfigure}
 \caption{Losses using phase-randomized states as a function of time $t$. The solid lines indicate $f_h$ for two values of $\mu=0.50$ (a) and $\mu=1.50$ (b), with parameters $\eta_m = 68\%$ and $\eta_d=100\%$, $95\%$ and $80\%$ from bottom to top. Symbols indicate the losses $f_d$ induced by the adversary to succeed with error rate $e=0$ or $1\%$. The protocol is secure as long as the solid line lies below the symbols.}
 \label{fig:memory}
 \end{figure}

In our security analysis, we have considered up till now fixed losses. However, in a quantum money implementation with a quantum memory used to store the credit card states, it will be necessary to take into account the time-dependent losses due to the decoherence of the memory. Here, the mint hands the stored quantum state to the client at time $t=0$. When $t>0$, the retrieval efficiency $\eta_{m}(t)$ decreases with time, thus increasing the losses to $e^{-\mu\eta_d\eta_{m}(t)}$. The initial retrieval efficiency $\eta_{m}(0)$ of the quantum memory limits the fraction of the $n$ states a dishonest client can retrieve to $\left(1-e^{-\mu\eta_d\eta_{m}(0)}\right)$. He may however have access to an ideal quantum memory to transfer the state at $t=0$, before $\eta_{m}(t)$ starts to decrease. This opens the door to powerful loss dependent attacks whose success increases as a function of time. As an illustration, we consider the cold atomic ensemble setup described in \cite{VHC:NC18}, with losses that are dominated by the dephasing of the collective atomic magnetic excitation with a lifetime $\tau \approx 15\us$ due to weak residual magnetic fields.
%
The retrieval efficiency decreases as $\eta_{m}(t) \approx \eta_{m}(0)e^{-t^2/\tau^2}$, where 
$\eta_{m}(0)\approx 68\%$ for this setup. Using this expression, we solve (\ref{bobby}) with phase-randomized states and derive secure credit card lifetimes of a few $\us$, as shown in Fig. \ref{fig:memory} for $\mu=0.50$ and $\mu=1.50$.


\section{Conclusion}

By establishing an optimization framework in the coherent state setting, we have derived secure regions of operation for quantum credit card schemes in both trusted and untrusted terminal scenarios. With phase-randomized states, we have shown that the former case can be secure using a setup with detection efficiency $\eta_d>80\%$ and noise tolerance around $e=1-2\%$, while the latter case requires tighter parameters: $\eta_d>95\%$ and noise tolerance lower than $e=1\%$. Using the duality of semidefinite programs, we have provided numerical evidence that the adversary cannot increase his/her cheating probability by correlating the $n$ states in the credit card. In such a setting, the uncertainty on the tolerated number of incorrect outcomes $en$ and excess losses $f_d\:n$ scales as $\sqrt{n}$.  We have finally provided a method to derive secure credit card lifetimes in the presence of a decohering quantum memory. This work encourages the future implementation of quantum credit card schemes with state-of-the-art quantum storage devices, as it provides a simple framework to derive practical security parameters in a semi-device-independent setting.

\section{Acknowledgments}

We thank Félix Hoffet and Julien Laurat for useful discussions about the modeling of the quantum memory, as well as Anthony Leverrier for his helpful comments on the manuscript. We acknowledge support from the European Union through the project ERC-2017-STG-758911 QUSCO, from the ANR through the project ANR-17-CE39-0005 quBIC, from BPI France through the project 143024 RISQ, and from Université Paris-Saclay through the Initiative de Recherche Interdisciplinaire.



\begin{thebibliography}{43}%
\makeatletter
\providecommand \@ifxundefined [1]{%
 \@ifx{#1\undefined}
}%
\providecommand \@ifnum [1]{%
 \ifnum #1\expandafter \@firstoftwo
 \else \expandafter \@secondoftwo
 \fi
}%
\providecommand \@ifx [1]{%
 \ifx #1\expandafter \@firstoftwo
 \else \expandafter \@secondoftwo
 \fi
}%
\providecommand \natexlab [1]{#1}%
\providecommand \enquote  [1]{``#1''}%
\providecommand \bibnamefont  [1]{#1}%
\providecommand \bibfnamefont [1]{#1}%
\providecommand \citenamefont [1]{#1}%
\providecommand \href@noop [0]{\@secondoftwo}%
\providecommand \href [0]{\begingroup \@sanitize@url \@href}%
\providecommand \@href[1]{\@@startlink{#1}\@@href}%
\providecommand \@@href[1]{\endgroup#1\@@endlink}%
\providecommand \@sanitize@url [0]{\catcode `\\12\catcode `\$12\catcode
  `\&12\catcode `\#12\catcode `\^12\catcode `\_12\catcode `\%12\relax}%
\providecommand \@@startlink[1]{}%
\providecommand \@@endlink[0]{}%
\providecommand \url  [0]{\begingroup\@sanitize@url \@url }%
\providecommand \@url [1]{\endgroup\@href {#1}{\urlprefix }}%
\providecommand \urlprefix  [0]{URL }%
\providecommand \Eprint [0]{\href }%
\providecommand \doibase [0]{http://dx.doi.org/}%
\providecommand \selectlanguage [0]{\@gobble}%
\providecommand \bibinfo  [0]{\@secondoftwo}%
\providecommand \bibfield  [0]{\@secondoftwo}%
\providecommand \translation [1]{[#1]}%
\providecommand \BibitemOpen [0]{}%
\providecommand \bibitemStop [0]{}%
\providecommand \bibitemNoStop [0]{.\EOS\space}%
\providecommand \EOS [0]{\spacefactor3000\relax}%
\providecommand \BibitemShut  [1]{\csname bibitem#1\endcsname}%
\let\auto@bib@innerbib\@empty
\bibitem [{\citenamefont {Wooters}\ and\ \citenamefont
  {Zurek}(1982)}]{WZ:nature82}%
  \BibitemOpen
  \bibfield  {author} {\bibinfo {author} {\bibfnamefont {W.~K.}\ \bibnamefont
  {Wooters}}\ and\ \bibinfo {author} {\bibfnamefont {W.~H.}\ \bibnamefont
  {Zurek}},\ }\href {\doibase 10.1038/299802a0} {\bibfield  {journal} {\bibinfo
   {journal} {Nature}\ }\textbf {\bibinfo {volume} {299}},\ \bibinfo {pages}
  {802} (\bibinfo {year} {1982})}\BibitemShut {NoStop}%
\bibitem [{\citenamefont {Wiesner}(1983)}]{Wie:acm83}%
  \BibitemOpen
  \bibfield  {author} {\bibinfo {author} {\bibfnamefont {S.}~\bibnamefont
  {Wiesner}},\ }\href {\doibase 10.1145/1008908.1008920} {\bibfield  {journal}
  {\bibinfo  {journal} {ACM Sigact News}\ }\textbf {\bibinfo {volume} {15}},\
  \bibinfo {pages} {78} (\bibinfo {year} {1983})}\BibitemShut {NoStop}%
\bibitem [{\citenamefont {Bennett}\ and\ \citenamefont
  {Brassard}(1984)}]{BB84}%
  \BibitemOpen
  \bibfield  {author} {\bibinfo {author} {\bibfnamefont {C.~H.}\ \bibnamefont
  {Bennett}}\ and\ \bibinfo {author} {\bibfnamefont {G.}~\bibnamefont
  {Brassard}},\ }in\ \href
  {https://researcher.watson.ibm.com/researcher/files/us-bennetc/BB84highest.pdf}
  {\emph {\bibinfo {booktitle} {Proc. IEEE International Conference on
  Computers, Systems and Signal Processing}}},\ Vol.~\bibinfo {volume} {1}\
  (\bibinfo {address} {Bangalore, India},\ \bibinfo {year} {1984})\ pp.\
  \bibinfo {pages} {175--179}\BibitemShut {NoStop}%
\bibitem [{\citenamefont {Gisin}\ \emph {et~al.}(2002)\citenamefont {Gisin},
  \citenamefont {Ribordy}, \citenamefont {Tittel},\ and\ \citenamefont
  {Zbinden}}]{GRT:RMP02}%
  \BibitemOpen
  \bibfield  {author} {\bibinfo {author} {\bibfnamefont {N.}~\bibnamefont
  {Gisin}}, \bibinfo {author} {\bibfnamefont {G.}~\bibnamefont {Ribordy}},
  \bibinfo {author} {\bibfnamefont {W.}~\bibnamefont {Tittel}}, \ and\ \bibinfo
  {author} {\bibfnamefont {H.}~\bibnamefont {Zbinden}},\ }\href {\doibase
  10.1103/RevModPhys.74.145} {\bibfield  {journal} {\bibinfo  {journal} {Rev.
  Mod. Phys.}\ }\textbf {\bibinfo {volume} {74}},\ \bibinfo {pages} {145}
  (\bibinfo {year} {2002})},\ \Eprint {http://arxiv.org/abs/quant-ph/0101098}
  {arXiv:quant-ph/0101098} \BibitemShut {NoStop}%
\bibitem [{\citenamefont {Scarani}\ \emph {et~al.}(2009)\citenamefont
  {Scarani}, \citenamefont {Bechmann-Pasquinucci}, \citenamefont {Cerf},
  \citenamefont {Du\v{s}ek}, \citenamefont {L\"utkenhaus},\ and\ \citenamefont
  {Peev}}]{SBP:RMP09}%
  \BibitemOpen
  \bibfield  {author} {\bibinfo {author} {\bibfnamefont {V.}~\bibnamefont
  {Scarani}}, \bibinfo {author} {\bibfnamefont {H.}~\bibnamefont
  {Bechmann-Pasquinucci}}, \bibinfo {author} {\bibfnamefont {N.}~\bibnamefont
  {Cerf}}, \bibinfo {author} {\bibfnamefont {M.}~\bibnamefont {Du\v{s}ek}},
  \bibinfo {author} {\bibfnamefont {N.}~\bibnamefont {L\"utkenhaus}}, \ and\
  \bibinfo {author} {\bibfnamefont {M.}~\bibnamefont {Peev}},\ }\href {\doibase
  10.1103/RevModPhys.81.1301} {\bibfield  {journal} {\bibinfo  {journal} {Rev.
  Mod. Phys.}\ }\textbf {\bibinfo {volume} {81}},\ \bibinfo {pages} {1301}
  (\bibinfo {year} {2009})},\ \Eprint {http://arxiv.org/abs/0802.4155}
  {arXiv:0802.4155} \BibitemShut {NoStop}%
\bibitem [{\citenamefont {Brodutch}\ \emph {et~al.}(2016)\citenamefont
  {Brodutch}, \citenamefont {Nagaj}, \citenamefont {Sattath},\ and\
  \citenamefont {Unruh}}]{BNS:QIC16}%
  \BibitemOpen
  \bibfield  {author} {\bibinfo {author} {\bibfnamefont {A.}~\bibnamefont
  {Brodutch}}, \bibinfo {author} {\bibfnamefont {D.}~\bibnamefont {Nagaj}},
  \bibinfo {author} {\bibfnamefont {O.}~\bibnamefont {Sattath}}, \ and\
  \bibinfo {author} {\bibfnamefont {D.}~\bibnamefont {Unruh}},\ }\href
  {\doibase 10.26421/QIC16.11-12} {\bibfield  {journal} {\bibinfo  {journal}
  {Quantum Information \& Computation}\ }\textbf {\bibinfo {volume} {16}},\
  \bibinfo {pages} {1048} (\bibinfo {year} {2016})},\ \Eprint
  {http://arxiv.org/abs/1404.1507} {arXiv:1404.1507} \BibitemShut {NoStop}%
\bibitem [{\citenamefont {Gavinski}(2012)}]{Gav:ccc12}%
  \BibitemOpen
  \bibfield  {author} {\bibinfo {author} {\bibfnamefont {D.}~\bibnamefont
  {Gavinski}},\ }in\ \href {\doibase 10.1109/CCC.2012.10} {\emph {\bibinfo
  {booktitle} {Proc. IEEE 27th Annual Conference on Computational Complexity
  (CCC)}}}\ (\bibinfo  {publisher} {IEEE},\ \bibinfo {year} {2012})\ pp.\
  \bibinfo {pages} {42--52},\ \Eprint {http://arxiv.org/abs/1109.0372}
  {arXiv:1109.0372} \BibitemShut {NoStop}%
\bibitem [{\citenamefont {Pastawski}\ \emph {et~al.}(2012)\citenamefont
  {Pastawski}, \citenamefont {Yao}, \citenamefont {Jiang}, \citenamefont
  {Lukin},\ and\ \citenamefont {Cirac}}]{PY+:pnas12}%
  \BibitemOpen
  \bibfield  {author} {\bibinfo {author} {\bibfnamefont {F.}~\bibnamefont
  {Pastawski}}, \bibinfo {author} {\bibfnamefont {N.~Y.}\ \bibnamefont {Yao}},
  \bibinfo {author} {\bibfnamefont {L.}~\bibnamefont {Jiang}}, \bibinfo
  {author} {\bibfnamefont {M.~D.}\ \bibnamefont {Lukin}}, \ and\ \bibinfo
  {author} {\bibfnamefont {J.~I.}\ \bibnamefont {Cirac}},\ }\href {\doibase
  10.1073/pnas.1203552109} {\bibfield  {journal} {\bibinfo  {journal} {PNAS}\
  }\textbf {\bibinfo {volume} {109}},\ \bibinfo {pages} {16079} (\bibinfo
  {year} {2012})},\ \Eprint {http://arxiv.org/abs/1112.5456} {arXiv:1112.5456}
  \BibitemShut {NoStop}%
\bibitem [{\citenamefont {Georgiou}\ and\ \citenamefont
  {Kerenidis}(2015)}]{GK:tqc15}%
  \BibitemOpen
  \bibfield  {author} {\bibinfo {author} {\bibfnamefont {M.}~\bibnamefont
  {Georgiou}}\ and\ \bibinfo {author} {\bibfnamefont {I.}~\bibnamefont
  {Kerenidis}},\ }in\ \href {\doibase 10.4230/LIPIcs.TQC.2015.92} {\emph
  {\bibinfo {booktitle} {Proc. 10th Conference on the Theory of Quantum
  Computation, Communication and Cryptography (TQC)}}},\ Vol.~\bibinfo {volume}
  {44},\ \bibinfo {editor} {edited by\ \bibinfo {editor} {\bibfnamefont
  {S.}~\bibnamefont {Beigi}}\ and\ \bibinfo {editor} {\bibfnamefont
  {R.}~\bibnamefont {K\"onig}}}\ (\bibinfo  {publisher} {Schlo\ss\
  Dagstuhl--Leibniz-Zentrum f\"ur Informatik},\ \bibinfo {address} {Dagstuhl,
  Germany},\ \bibinfo {year} {2015})\ pp.\ \bibinfo {pages}
  {92--110}\BibitemShut {NoStop}%
\bibitem [{\citenamefont {Amiri}\ and\ \citenamefont
  {Arrazola}(2017)}]{AA:pra17}%
  \BibitemOpen
  \bibfield  {author} {\bibinfo {author} {\bibfnamefont {R.}~\bibnamefont
  {Amiri}}\ and\ \bibinfo {author} {\bibfnamefont {J.~M.}\ \bibnamefont
  {Arrazola}},\ }\href {\doibase 10.1103/PhysRevA.95.062334} {\bibfield
  {journal} {\bibinfo  {journal} {Phys. Rev. A}\ }\textbf {\bibinfo {volume}
  {95}},\ \bibinfo {pages} {062334} (\bibinfo {year} {2017})},\ \Eprint
  {http://arxiv.org/abs/1610.06345} {arXiv:1610.06345} \BibitemShut {NoStop}%
\bibitem [{\citenamefont {Diamanti}\ \emph {et~al.}(2016)\citenamefont
  {Diamanti}, \citenamefont {Lo}, \citenamefont {Qi},\ and\ \citenamefont
  {Yuan}}]{DL+:npjQI16}%
  \BibitemOpen
  \bibfield  {author} {\bibinfo {author} {\bibfnamefont {E.}~\bibnamefont
  {Diamanti}}, \bibinfo {author} {\bibfnamefont {H.-K.}\ \bibnamefont {Lo}},
  \bibinfo {author} {\bibfnamefont {B.}~\bibnamefont {Qi}}, \ and\ \bibinfo
  {author} {\bibfnamefont {Z.}~\bibnamefont {Yuan}},\ }\href {\doibase
  10.1038/npjqi.2016.25} {\bibfield  {journal} {\bibinfo  {journal} {npj
  Quantum Information}\ }\textbf {\bibinfo {volume} {2}},\ \bibinfo {pages}
  {16025} (\bibinfo {year} {2016})},\ \Eprint {http://arxiv.org/abs/1606.05853}
  {arXiv:1606.05853} \BibitemShut {NoStop}%
\bibitem [{\citenamefont {Heshami}\ \emph {et~al.}(2016)\citenamefont
  {Heshami}, \citenamefont {England}, \citenamefont {Humphreys}, \citenamefont
  {Bustard}, \citenamefont {Acosta}, \citenamefont {Nunn},\ and\ \citenamefont
  {Sussman}}]{HE+:jmo16}%
  \BibitemOpen
  \bibfield  {author} {\bibinfo {author} {\bibfnamefont {K.}~\bibnamefont
  {Heshami}}, \bibinfo {author} {\bibfnamefont {D.~G.}\ \bibnamefont
  {England}}, \bibinfo {author} {\bibfnamefont {P.~C.}\ \bibnamefont
  {Humphreys}}, \bibinfo {author} {\bibfnamefont {P.~J.}\ \bibnamefont
  {Bustard}}, \bibinfo {author} {\bibfnamefont {V.~M.}\ \bibnamefont {Acosta}},
  \bibinfo {author} {\bibfnamefont {J.}~\bibnamefont {Nunn}}, \ and\ \bibinfo
  {author} {\bibfnamefont {B.~J.}\ \bibnamefont {Sussman}},\ }\href {\doibase
  10.1080/09500340.2016.1148212} {\bibfield  {journal} {\bibinfo  {journal} {J.
  Mod. Opt.}\ }\textbf {\bibinfo {volume} {63}},\ \bibinfo {pages} {2005}
  (\bibinfo {year} {2016})},\ \Eprint {http://arxiv.org/abs/1511.04018}
  {arXiv:1511.04018} \BibitemShut {NoStop}%
\bibitem [{\citenamefont {Bartkiewicz}\ \emph {et~al.}(2017)\citenamefont
  {Bartkiewicz}, \citenamefont {\v{C}ernoch}, \citenamefont {Chimczak},
  \citenamefont {Lemr}, \citenamefont {Miranowicz},\ and\ \citenamefont
  {Nori}}]{BC+:npjQI17}%
  \BibitemOpen
  \bibfield  {author} {\bibinfo {author} {\bibfnamefont {K.}~\bibnamefont
  {Bartkiewicz}}, \bibinfo {author} {\bibfnamefont {A.}~\bibnamefont
  {\v{C}ernoch}}, \bibinfo {author} {\bibfnamefont {G.}~\bibnamefont
  {Chimczak}}, \bibinfo {author} {\bibfnamefont {K.}~\bibnamefont {Lemr}},
  \bibinfo {author} {\bibfnamefont {A.}~\bibnamefont {Miranowicz}}, \ and\
  \bibinfo {author} {\bibfnamefont {F.}~\bibnamefont {Nori}},\ }\href {\doibase
  10.1038/s41534-017-0010-x} {\bibfield  {journal} {\bibinfo  {journal} {npj
  Quantum Information}\ }\textbf {\bibinfo {volume} {3}},\ \bibinfo {pages} {7}
  (\bibinfo {year} {2017})},\ \Eprint {http://arxiv.org/abs/1604.04453}
  {arXiv:1604.04453} \BibitemShut {NoStop}%
\bibitem [{\citenamefont {Bozzio}\ \emph {et~al.}(2018)\citenamefont {Bozzio},
  \citenamefont {Orieux}, \citenamefont {Vidarte}, \citenamefont {Zaquine},
  \citenamefont {Kerenidis},\ and\ \citenamefont {Diamanti}}]{BOV:npj18}%
  \BibitemOpen
  \bibfield  {author} {\bibinfo {author} {\bibfnamefont {M.}~\bibnamefont
  {Bozzio}}, \bibinfo {author} {\bibfnamefont {A.}~\bibnamefont {Orieux}},
  \bibinfo {author} {\bibfnamefont {L.~T.}\ \bibnamefont {Vidarte}}, \bibinfo
  {author} {\bibfnamefont {I.}~\bibnamefont {Zaquine}}, \bibinfo {author}
  {\bibfnamefont {I.}~\bibnamefont {Kerenidis}}, \ and\ \bibinfo {author}
  {\bibfnamefont {E.}~\bibnamefont {Diamanti}},\ }\href {\doibase
  10.1038/s41534-018-0058-2} {\bibfield  {journal} {\bibinfo  {journal} {npj
  Quantum Information}\ }\textbf {\bibinfo {volume} {4}},\ \bibinfo {pages} {5}
  (\bibinfo {year} {2018})},\ \Eprint {http://arxiv.org/abs/1705.01428}
  {arXiv:1705.01428} \BibitemShut {NoStop}%
\bibitem [{\citenamefont {Guan}\ \emph {et~al.}(2018)\citenamefont {Guan},
  \citenamefont {Arrazola}, \citenamefont {Amiri}, \citenamefont {Zhang},
  \citenamefont {Li}, \citenamefont {You}, \citenamefont {Wang}, \citenamefont
  {Zhang},\ and\ \citenamefont {Pan}}]{GAA:pra18}%
  \BibitemOpen
  \bibfield  {author} {\bibinfo {author} {\bibfnamefont {J.-Y.}\ \bibnamefont
  {Guan}}, \bibinfo {author} {\bibfnamefont {J.-M.}\ \bibnamefont {Arrazola}},
  \bibinfo {author} {\bibfnamefont {R.}~\bibnamefont {Amiri}}, \bibinfo
  {author} {\bibfnamefont {W.}~\bibnamefont {Zhang}}, \bibinfo {author}
  {\bibfnamefont {H.}~\bibnamefont {Li}}, \bibinfo {author} {\bibfnamefont
  {L.}~\bibnamefont {You}}, \bibinfo {author} {\bibfnamefont {Z.}~\bibnamefont
  {Wang}}, \bibinfo {author} {\bibfnamefont {Q.}~\bibnamefont {Zhang}}, \ and\
  \bibinfo {author} {\bibfnamefont {J.-W.}\ \bibnamefont {Pan}},\ }\href
  {\doibase 10.1103/PhysRevA.97.032338} {\bibfield  {journal} {\bibinfo
  {journal} {Phys. Rev. A}\ }\textbf {\bibinfo {volume} {97}},\ \bibinfo
  {pages} {032338} (\bibinfo {year} {2018})},\ \Eprint
  {http://arxiv.org/abs/1709.05882} {arXiv:1709.05882} \BibitemShut {NoStop}%
\bibitem [{\citenamefont {Acin}\ \emph {et~al.}(2007)\citenamefont {Acin},
  \citenamefont {Brunner}, \citenamefont {Gisin}, \citenamefont {Massar},
  \citenamefont {Pironio},\ and\ \citenamefont {Scarani}}]{ABG:PRL07}%
  \BibitemOpen
  \bibfield  {author} {\bibinfo {author} {\bibfnamefont {A.}~\bibnamefont
  {Acin}}, \bibinfo {author} {\bibfnamefont {N.}~\bibnamefont {Brunner}},
  \bibinfo {author} {\bibfnamefont {N.}~\bibnamefont {Gisin}}, \bibinfo
  {author} {\bibfnamefont {S.}~\bibnamefont {Massar}}, \bibinfo {author}
  {\bibfnamefont {S.}~\bibnamefont {Pironio}}, \ and\ \bibinfo {author}
  {\bibfnamefont {V.}~\bibnamefont {Scarani}},\ }\href {\doibase
  10.1103/PhysRevLett.98.230501} {\bibfield  {journal} {\bibinfo  {journal}
  {Phys. Rev. Lett.}\ }\textbf {\bibinfo {volume} {98}},\ \bibinfo {pages}
  {230501} (\bibinfo {year} {2007})},\ \Eprint {http://arxiv.org/abs/0702152}
  {arXiv:0702152} \BibitemShut {NoStop}%
\bibitem [{\citenamefont {Ma}\ and\ \citenamefont
  {L\"utkenhaus}(2012)}]{ML:QIC12}%
  \BibitemOpen
  \bibfield  {author} {\bibinfo {author} {\bibfnamefont {X.}~\bibnamefont
  {Ma}}\ and\ \bibinfo {author} {\bibfnamefont {N.}~\bibnamefont
  {L\"utkenhaus}},\ }\href {\doibase 10.26421/QIC12.3-4} {\bibfield  {journal}
  {\bibinfo  {journal} {Quantum Information \& Computation}\ }\textbf {\bibinfo
  {volume} {12}},\ \bibinfo {pages} {202} (\bibinfo {year} {2012})},\ \Eprint
  {http://arxiv.org/abs/1109.1203} {arXiv:1109.1203} \BibitemShut {NoStop}%
\bibitem [{\citenamefont {Branciard}\ \emph {et~al.}(2012)\citenamefont
  {Branciard}, \citenamefont {Cavalcanti}, \citenamefont {Walborn},
  \citenamefont {Scarani},\ and\ \citenamefont {Wiseman}}]{BCW:PRA12}%
  \BibitemOpen
  \bibfield  {author} {\bibinfo {author} {\bibfnamefont {C.}~\bibnamefont
  {Branciard}}, \bibinfo {author} {\bibfnamefont {E.~G.}\ \bibnamefont
  {Cavalcanti}}, \bibinfo {author} {\bibfnamefont {S.~P.}\ \bibnamefont
  {Walborn}}, \bibinfo {author} {\bibfnamefont {V.}~\bibnamefont {Scarani}}, \
  and\ \bibinfo {author} {\bibfnamefont {H.~M.}\ \bibnamefont {Wiseman}},\
  }\href {\doibase 10.1103/PhysRevA.85.010301} {\bibfield  {journal} {\bibinfo
  {journal} {Phys. Rev. A}\ }\textbf {\bibinfo {volume} {85}},\ \bibinfo
  {pages} {010301} (\bibinfo {year} {2012})},\ \Eprint
  {http://arxiv.org/abs/1109.1435} {arXiv:1109.1435} \BibitemShut {NoStop}%
\bibitem [{\citenamefont {Lo}\ \emph {et~al.}(2012)\citenamefont {Lo},
  \citenamefont {Curty},\ and\ \citenamefont {Qi}}]{LCQ:PRL12}%
  \BibitemOpen
  \bibfield  {author} {\bibinfo {author} {\bibfnamefont {H.-K.}\ \bibnamefont
  {Lo}}, \bibinfo {author} {\bibfnamefont {M.}~\bibnamefont {Curty}}, \ and\
  \bibinfo {author} {\bibfnamefont {B.}~\bibnamefont {Qi}},\ }\href {\doibase
  10.1103/PhysRevLett.108.130503} {\bibfield  {journal} {\bibinfo  {journal}
  {Phys. Rev. Lett.}\ }\textbf {\bibinfo {volume} {108}},\ \bibinfo {pages}
  {130503} (\bibinfo {year} {2012})},\ \Eprint {http://arxiv.org/abs/1109.1473}
  {arXiv:1109.1473} \BibitemShut {NoStop}%
\bibitem [{\citenamefont {Braunstein}\ and\ \citenamefont
  {Pirandola}(2012)}]{BP:PRL12}%
  \BibitemOpen
  \bibfield  {author} {\bibinfo {author} {\bibfnamefont {S.~L.}\ \bibnamefont
  {Braunstein}}\ and\ \bibinfo {author} {\bibfnamefont {S.}~\bibnamefont
  {Pirandola}},\ }\href {\doibase 10.1103/PhysRevLett.108.130502} {\bibfield
  {journal} {\bibinfo  {journal} {Phys. Rev. Lett.}\ }\textbf {\bibinfo
  {volume} {108}},\ \bibinfo {pages} {130502} (\bibinfo {year} {2012})},\
  \Eprint {http://arxiv.org/abs/1109.2330} {arXiv:1109.2330} \BibitemShut
  {NoStop}%
\bibitem [{\citenamefont {Liang}\ \emph {et~al.}(2011)\citenamefont {Liang},
  \citenamefont {V\'ertesi},\ and\ \citenamefont {Brunner}}]{LVB:PRA11}%
  \BibitemOpen
  \bibfield  {author} {\bibinfo {author} {\bibfnamefont {Y.-C.}\ \bibnamefont
  {Liang}}, \bibinfo {author} {\bibfnamefont {T.}~\bibnamefont {V\'ertesi}}, \
  and\ \bibinfo {author} {\bibfnamefont {N.}~\bibnamefont {Brunner}},\ }\href
  {\doibase 10.1103/PhysRevA.83.022108} {\bibfield  {journal} {\bibinfo
  {journal} {Phys. Rev. A}\ }\textbf {\bibinfo {volume} {83}},\ \bibinfo
  {pages} {022108} (\bibinfo {year} {2011})},\ \Eprint
  {http://arxiv.org/abs/1012.1513} {arXiv:1012.1513} \BibitemShut {NoStop}%
\bibitem [{\citenamefont {Paw\l{}owski}\ and\ \citenamefont
  {Brunner}(2011)}]{PB:PRA11}%
  \BibitemOpen
  \bibfield  {author} {\bibinfo {author} {\bibfnamefont {M.}~\bibnamefont
  {Paw\l{}owski}}\ and\ \bibinfo {author} {\bibfnamefont {N.}~\bibnamefont
  {Brunner}},\ }\href {\doibase 10.1103/PhysRevA.84.010302} {\bibfield
  {journal} {\bibinfo  {journal} {Phys. Rev. A}\ }\textbf {\bibinfo {volume}
  {84}},\ \bibinfo {pages} {010302} (\bibinfo {year} {2011})},\ \Eprint
  {http://arxiv.org/abs/1103.4105} {arXiv:1103.4105} \BibitemShut {NoStop}%
\bibitem [{\citenamefont {Gittsovich}\ and\ \citenamefont
  {Moroder}(2013)}]{CM:AIP13}%
  \BibitemOpen
  \bibfield  {author} {\bibinfo {author} {\bibfnamefont {O.}~\bibnamefont
  {Gittsovich}}\ and\ \bibinfo {author} {\bibfnamefont {T.}~\bibnamefont
  {Moroder}},\ }in\ \href {\doibase 10.1063/1.4903122} {\emph {\bibinfo
  {booktitle} {Proc. of 11th Int. Conf. on Quantum Communication, Measurement
  and Computation (QCMC12)}}},\ \bibinfo {series} {AIP Conference Proceedings},
  Vol.\ \bibinfo {volume} {1633}\ (\bibinfo  {publisher} {American Institute of
  Physics},\ \bibinfo {year} {2013})\ p.\ \bibinfo {pages} {156},\ \Eprint
  {http://arxiv.org/abs/1303.3484} {arXiv:1303.3484} \BibitemShut {NoStop}%
\bibitem [{\citenamefont {Van~Himbeeck}\ \emph {et~al.}(2017)\citenamefont
  {Van~Himbeeck}, \citenamefont {Woodhead}, \citenamefont {Cerf}, \citenamefont
  {Garc\'ia-Patr\'on},\ and\ \citenamefont {Pironio}}]{VHWC:Qua17}%
  \BibitemOpen
  \bibfield  {author} {\bibinfo {author} {\bibfnamefont {T.}~\bibnamefont
  {Van~Himbeeck}}, \bibinfo {author} {\bibfnamefont {E.}~\bibnamefont
  {Woodhead}}, \bibinfo {author} {\bibfnamefont {N.~J.}\ \bibnamefont {Cerf}},
  \bibinfo {author} {\bibfnamefont {R.}~\bibnamefont {Garc\'ia-Patr\'on}}, \
  and\ \bibinfo {author} {\bibfnamefont {S.}~\bibnamefont {Pironio}},\ }\href
  {\doibase 10.22331/q-2017-11-18-33} {\bibfield  {journal} {\bibinfo
  {journal} {Quantum}\ }\textbf {\bibinfo {volume} {1}},\ \bibinfo {pages} {33}
  (\bibinfo {year} {2017})},\ \Eprint {http://arxiv.org/abs/1612.06828}
  {arXiv:1612.06828} \BibitemShut {NoStop}%
\bibitem [{\citenamefont {Horodecki}\ and\ \citenamefont
  {Stankiewicz}(2018)}]{HS:arx18}%
  \BibitemOpen
  \bibfield  {author} {\bibinfo {author} {\bibfnamefont {K.}~\bibnamefont
  {Horodecki}}\ and\ \bibinfo {author} {\bibfnamefont {M.}~\bibnamefont
  {Stankiewicz}},\ }\href@noop {} {\  (\bibinfo {year} {2018})},\ \Eprint
  {http://arxiv.org/abs/1811.10552} {arXiv:1811.10552} \BibitemShut {NoStop}%
\bibitem [{\citenamefont {Jir\'akov\'a}\ \emph {et~al.}(2018)\citenamefont
  {Jir\'akov\'a}, \citenamefont {Bartkiewicz}, \citenamefont {\v{C}ernoch},\
  and\ \citenamefont {Lemr}}]{JBC:arx18}%
  \BibitemOpen
  \bibfield  {author} {\bibinfo {author} {\bibfnamefont {K.}~\bibnamefont
  {Jir\'akov\'a}}, \bibinfo {author} {\bibfnamefont {K.}~\bibnamefont
  {Bartkiewicz}}, \bibinfo {author} {\bibfnamefont {A.}~\bibnamefont
  {\v{C}ernoch}}, \ and\ \bibinfo {author} {\bibfnamefont {K.}~\bibnamefont
  {Lemr}},\ }\href@noop {} {\  (\bibinfo {year} {2018})},\ \Eprint
  {http://arxiv.org/abs/1811.10718} {arXiv:1811.10718} \BibitemShut {NoStop}%
\bibitem [{\citenamefont {Molina}\ \emph {et~al.}(2013)\citenamefont {Molina},
  \citenamefont {Vidick},\ and\ \citenamefont {Watrous}}]{MVW:tqc12}%
  \BibitemOpen
  \bibfield  {author} {\bibinfo {author} {\bibfnamefont {A.}~\bibnamefont
  {Molina}}, \bibinfo {author} {\bibfnamefont {T.}~\bibnamefont {Vidick}}, \
  and\ \bibinfo {author} {\bibfnamefont {J.}~\bibnamefont {Watrous}},\ }in\
  \href {\doibase 10.1007/978-3-642-35656-8_4} {\emph {\bibinfo {booktitle}
  {{TQC} 2012: Theory of Quantum Computation, Communication, and
  Cryptography}}},\ \bibinfo {series} {Lecture Notes in Computer Science},
  Vol.\ \bibinfo {volume} {7582},\ \bibinfo {editor} {edited by\ \bibinfo
  {editor} {\bibfnamefont {K.}~\bibnamefont {Iwama}}, \bibinfo {editor}
  {\bibfnamefont {Y.}~\bibnamefont {Kawano}}, \ and\ \bibinfo {editor}
  {\bibfnamefont {M.}~\bibnamefont {Murao}}}\ (\bibinfo  {publisher}
  {Springer},\ \bibinfo {year} {2013})\ \Eprint
  {http://arxiv.org/abs/1202.4010} {arXiv:1202.4010} \BibitemShut {NoStop}%
\bibitem [{\citenamefont {Watrous}(2011)}]{W:LN11}%
  \BibitemOpen
  \bibfield  {author} {\bibinfo {author} {\bibfnamefont {J.}~\bibnamefont
  {Watrous}},\ }\enquote {\bibinfo {title} {Semidefinite programming},}\ in\
  \href {https://cs.uwaterloo.ca/~watrous/LectureNotes.html} {\emph {\bibinfo
  {booktitle} {Theory of Quantum Information (notes from Fall 2011)}}}\
  (\bibinfo  {publisher} {University of Waterloo},\ \bibinfo {year} {2011})\
  Chap.~\bibinfo {chapter} {7}\BibitemShut {NoStop}%
\bibitem [{\citenamefont {Vandenberghe}\ and\ \citenamefont
  {Boyd}(1996)}]{VB:SIAM96}%
  \BibitemOpen
  \bibfield  {author} {\bibinfo {author} {\bibfnamefont {L.}~\bibnamefont
  {Vandenberghe}}\ and\ \bibinfo {author} {\bibfnamefont {S.}~\bibnamefont
  {Boyd}},\ }\href {\doibase 10.1137/1038003} {\bibfield  {journal} {\bibinfo
  {journal} {SIAM Review}\ }\textbf {\bibinfo {volume} {38}},\ \bibinfo {pages}
  {49} (\bibinfo {year} {1996})}\BibitemShut {NoStop}%
\bibitem [{\citenamefont {Beaudry}\ \emph {et~al.}(2008)\citenamefont
  {Beaudry}, \citenamefont {Moroder},\ and\ \citenamefont
  {L\"utkenhaus}}]{BML:prl08}%
  \BibitemOpen
  \bibfield  {author} {\bibinfo {author} {\bibfnamefont {N.~J.}\ \bibnamefont
  {Beaudry}}, \bibinfo {author} {\bibfnamefont {T.}~\bibnamefont {Moroder}}, \
  and\ \bibinfo {author} {\bibfnamefont {N.}~\bibnamefont {L\"utkenhaus}},\
  }\href {\doibase 10.1103/PhysRevLett.101.093601} {\bibfield  {journal}
  {\bibinfo  {journal} {Phys. Rev. Lett.}\ }\textbf {\bibinfo {volume} {101}},\
  \bibinfo {pages} {093601} (\bibinfo {year} {2008})},\ \Eprint
  {http://arxiv.org/abs/0804.3082} {arXiv:0804.3082} \BibitemShut {NoStop}%
\bibitem [{\citenamefont {Gittsovich}\ \emph {et~al.}(2014)\citenamefont
  {Gittsovich}, \citenamefont {Beaudry}, \citenamefont {Narasimhachar},
  \citenamefont {Alvarez}, \citenamefont {Moroder},\ and\ \citenamefont
  {L\"utkenhaus}}]{GBN:pra14}%
  \BibitemOpen
  \bibfield  {author} {\bibinfo {author} {\bibfnamefont {O.}~\bibnamefont
  {Gittsovich}}, \bibinfo {author} {\bibfnamefont {N.}~\bibnamefont {Beaudry}},
  \bibinfo {author} {\bibfnamefont {V.}~\bibnamefont {Narasimhachar}}, \bibinfo
  {author} {\bibfnamefont {R.~R.}\ \bibnamefont {Alvarez}}, \bibinfo {author}
  {\bibfnamefont {T.}~\bibnamefont {Moroder}}, \ and\ \bibinfo {author}
  {\bibfnamefont {N.}~\bibnamefont {L\"utkenhaus}},\ }\href {\doibase
  10.1103/PhysRevA.89.012325} {\bibfield  {journal} {\bibinfo  {journal} {Phys.
  Rev. A}\ }\textbf {\bibinfo {volume} {89}},\ \bibinfo {pages} {012325}
  (\bibinfo {year} {2014})},\ \Eprint {http://arxiv.org/abs/1310.5059}
  {arXiv:1310.5059} \BibitemShut {NoStop}%
\bibitem [{\citenamefont {Lo}\ and\ \citenamefont {Preskill}(2007)}]{LP:QIC07}%
  \BibitemOpen
  \bibfield  {author} {\bibinfo {author} {\bibfnamefont {H.-K.}\ \bibnamefont
  {Lo}}\ and\ \bibinfo {author} {\bibfnamefont {J.}~\bibnamefont {Preskill}},\
  }\href@noop {} {\bibfield  {journal} {\bibinfo  {journal} {Quant. Inf.
  Comput.}\ }\textbf {\bibinfo {volume} {8}},\ \bibinfo {pages} {431} (\bibinfo
  {year} {2007})},\ \Eprint {http://arxiv.org/abs/0610203} {arXiv:0610203}
  \BibitemShut {NoStop}%
\bibitem [{\citenamefont {Toh}\ \emph {et~al.}(1999)\citenamefont {Toh},
  \citenamefont {Todd},\ and\ \citenamefont {T{\"u}t{\"u}nc{\"u}}}]{SDPT3:a}%
  \BibitemOpen
  \bibfield  {author} {\bibinfo {author} {\bibfnamefont {K.~C.}\ \bibnamefont
  {Toh}}, \bibinfo {author} {\bibfnamefont {M.~J.}\ \bibnamefont {Todd}}, \
  and\ \bibinfo {author} {\bibfnamefont {R.~H.}\ \bibnamefont
  {T{\"u}t{\"u}nc{\"u}}},\ }\href {\doibase 10.1080/10556789908805762}
  {\bibfield  {journal} {\bibinfo  {journal} {Optimization Methods and
  Software}\ }\textbf {\bibinfo {volume} {11}},\ \bibinfo {pages} {545}
  (\bibinfo {year} {1999})}\BibitemShut {NoStop}%
\bibitem [{\citenamefont {T{\"u}t{\"u}nc{\"u}}\ \emph
  {et~al.}(2003)\citenamefont {T{\"u}t{\"u}nc{\"u}}, \citenamefont {Toh},\ and\
  \citenamefont {Todd}}]{SDPT3:b}%
  \BibitemOpen
  \bibfield  {author} {\bibinfo {author} {\bibfnamefont {R.~H.}\ \bibnamefont
  {T{\"u}t{\"u}nc{\"u}}}, \bibinfo {author} {\bibfnamefont {K.~C.}\
  \bibnamefont {Toh}}, \ and\ \bibinfo {author} {\bibfnamefont {M.~J.}\
  \bibnamefont {Todd}},\ }\href {\doibase 10.1007/s10107-002-0347-5} {\bibfield
   {journal} {\bibinfo  {journal} {Mathematical Programming}\ }\textbf
  {\bibinfo {volume} {95}},\ \bibinfo {pages} {189} (\bibinfo {year}
  {2003})}\BibitemShut {NoStop}%
\bibitem [{\citenamefont {Grant}\ and\ \citenamefont {Boyd}(2014)}]{cvx}%
  \BibitemOpen
  \bibfield  {author} {\bibinfo {author} {\bibfnamefont {M.}~\bibnamefont
  {Grant}}\ and\ \bibinfo {author} {\bibfnamefont {S.}~\bibnamefont {Boyd}},\
  }\href@noop {} {\enquote {\bibinfo {title} {{CVX}: Matlab software for
  disciplined convex programming, version 2.1},}\ }\bibinfo {howpublished}
  {\url{http://cvxr.com/cvx}} (\bibinfo {year} {2014})\BibitemShut {NoStop}%
\bibitem [{\citenamefont {Grant}\ and\ \citenamefont {Boyd}(2008)}]{gb08}%
  \BibitemOpen
  \bibfield  {author} {\bibinfo {author} {\bibfnamefont {M.}~\bibnamefont
  {Grant}}\ and\ \bibinfo {author} {\bibfnamefont {S.}~\bibnamefont {Boyd}},\
  }in\ \href {\doibase 10.1007/978-1-84800-155-8_7} {\emph {\bibinfo
  {booktitle} {Recent Advances in Learning and Control}}},\ \bibinfo {series}
  {Lecture Notes in Control and Information Sciences}, Vol.\ \bibinfo {volume}
  {371},\ \bibinfo {editor} {edited by\ \bibinfo {editor} {\bibfnamefont
  {V.}~\bibnamefont {Blondel}}, \bibinfo {editor} {\bibfnamefont
  {S.}~\bibnamefont {Boyd}}, \ and\ \bibinfo {editor} {\bibfnamefont
  {H.}~\bibnamefont {Kimura}}}\ (\bibinfo  {publisher} {Springer},\ \bibinfo
  {year} {2008})\ pp.\ \bibinfo {pages} {95--110}\BibitemShut {NoStop}%
\bibitem [{\citenamefont {Lo}\ and\ \citenamefont
  {Preskill}(2005)}]{LP:calt05}%
  \BibitemOpen
  \bibfield  {author} {\bibinfo {author} {\bibfnamefont {H.-K.}\ \bibnamefont
  {Lo}}\ and\ \bibinfo {author} {\bibfnamefont {J.}~\bibnamefont {Preskill}},\
  }\href@noop {} {\  (\bibinfo {year} {2005})},\ \Eprint
  {http://arxiv.org/abs/quant-ph/0504209} {arXiv:quant-ph/0504209} \BibitemShut
  {NoStop}%
\bibitem [{\citenamefont {Zhao}\ \emph {et~al.}(2007)\citenamefont {Zhao},
  \citenamefont {Qi},\ and\ \citenamefont {Lo}}]{ZQL:apl07}%
  \BibitemOpen
  \bibfield  {author} {\bibinfo {author} {\bibfnamefont {Y.}~\bibnamefont
  {Zhao}}, \bibinfo {author} {\bibfnamefont {B.}~\bibnamefont {Qi}}, \ and\
  \bibinfo {author} {\bibfnamefont {H.-K.}\ \bibnamefont {Lo}},\ }\href
  {\doibase 10.1063/1.2432296} {\bibfield  {journal} {\bibinfo  {journal}
  {Appl. Phys. Lett.}\ }\textbf {\bibinfo {volume} {90}},\ \bibinfo {pages}
  {044106} (\bibinfo {year} {2007})}\BibitemShut {NoStop}%
\bibitem [{\citenamefont {Yuan}\ \emph {et~al.}(2014)\citenamefont {Yuan},
  \citenamefont {Lucamarini}, \citenamefont {Dynes}, \citenamefont
  {Fr\"ohlich}, \citenamefont {Ward},\ and\ \citenamefont
  {Shields}}]{YLD:PRAPP14}%
  \BibitemOpen
  \bibfield  {author} {\bibinfo {author} {\bibfnamefont {Z.~L.}\ \bibnamefont
  {Yuan}}, \bibinfo {author} {\bibfnamefont {M.}~\bibnamefont {Lucamarini}},
  \bibinfo {author} {\bibfnamefont {J.~F.}\ \bibnamefont {Dynes}}, \bibinfo
  {author} {\bibfnamefont {B.}~\bibnamefont {Fr\"ohlich}}, \bibinfo {author}
  {\bibfnamefont {M.~B.}\ \bibnamefont {Ward}}, \ and\ \bibinfo {author}
  {\bibfnamefont {A.~J.}\ \bibnamefont {Shields}},\ }\href {\doibase
  10.1103/PhysRevApplied.2.064006} {\bibfield  {journal} {\bibinfo  {journal}
  {Phys. Rev. Applied}\ }\textbf {\bibinfo {volume} {2}},\ \bibinfo {pages}
  {064006} (\bibinfo {year} {2014})},\ \Eprint
  {http://arxiv.org/abs/1501.01900} {arXiv:1501.01900} \BibitemShut {NoStop}%
\bibitem [{\citenamefont {Hadfield}(2009)}]{H:NP09}%
  \BibitemOpen
  \bibfield  {author} {\bibinfo {author} {\bibfnamefont {R.~H.}\ \bibnamefont
  {Hadfield}},\ }\href {\doibase 10.1038/nphoton.2009.230} {\bibfield
  {journal} {\bibinfo  {journal} {Nat. Photonics}\ }\textbf {\bibinfo {volume}
  {3}},\ \bibinfo {pages} {696} (\bibinfo {year} {2009})}\BibitemShut {NoStop}%
\bibitem [{\citenamefont {Lita}\ \emph {et~al.}(2008)\citenamefont {Lita},
  \citenamefont {Miller},\ and\ \citenamefont {Woo~Nam}}]{LMN:OE08}%
  \BibitemOpen
  \bibfield  {author} {\bibinfo {author} {\bibfnamefont {A.~E.}\ \bibnamefont
  {Lita}}, \bibinfo {author} {\bibfnamefont {A.~J.}\ \bibnamefont {Miller}}, \
  and\ \bibinfo {author} {\bibfnamefont {S.}~\bibnamefont {Woo~Nam}},\ }\href
  {\doibase 10.1364/OE.16.003032} {\bibfield  {journal} {\bibinfo  {journal}
  {Opt. Express}\ }\textbf {\bibinfo {volume} {16}},\ \bibinfo {pages} {3032}
  (\bibinfo {year} {2008})}\BibitemShut {NoStop}%
\bibitem [{\citenamefont {Vernaz-Gris}\ \emph {et~al.}(2018)\citenamefont
  {Vernaz-Gris}, \citenamefont {Huang}, \citenamefont {Cao}, \citenamefont
  {Sheremet},\ and\ \citenamefont {Laurat}}]{VHC:NC18}%
  \BibitemOpen
  \bibfield  {author} {\bibinfo {author} {\bibfnamefont {P.}~\bibnamefont
  {Vernaz-Gris}}, \bibinfo {author} {\bibfnamefont {K.}~\bibnamefont {Huang}},
  \bibinfo {author} {\bibfnamefont {M.}~\bibnamefont {Cao}}, \bibinfo {author}
  {\bibfnamefont {A.}~\bibnamefont {Sheremet}}, \ and\ \bibinfo {author}
  {\bibfnamefont {J.}~\bibnamefont {Laurat}},\ }\href {\doibase
  10.1038/s41467-017-02775-8} {\bibfield  {journal} {\bibinfo  {journal} {Nat.
  Comm.}\ }\textbf {\bibinfo {volume} {9}},\ \bibinfo {pages} {363} (\bibinfo
  {year} {2018})},\ \Eprint {http://arxiv.org/abs/1707.09372}
  {arXiv:1707.09372} \BibitemShut {NoStop}%
\bibitem [{\citenamefont {D'Ariano}\ \emph {et~al.}(2017)\citenamefont
  {D'Ariano}, \citenamefont {Chiribella},\ and\ \citenamefont
  {Perinotti}}]{DCP:CUP17}%
  \BibitemOpen
  \bibfield  {author} {\bibinfo {author} {\bibfnamefont {G.~M.}\ \bibnamefont
  {D'Ariano}}, \bibinfo {author} {\bibfnamefont {G.}~\bibnamefont
  {Chiribella}}, \ and\ \bibinfo {author} {\bibfnamefont {P.}~\bibnamefont
  {Perinotti}},\ }in\ \href {\doibase 10.1017/9781107338340} {\emph {\bibinfo
  {booktitle} {Quantum Theory from First Principles, An Informational
  Approach}}}\ (\bibinfo  {publisher} {Cambridge University Press},\ \bibinfo
  {year} {2017})\ pp.\ \bibinfo {pages} {39--40}\BibitemShut {NoStop}%
\end{thebibliography}

%


\section*{Appendix A : Choi--Jamio\l kowski operator and semidefinite programming.}

\subsection*{A.1 : Choi's theorem on completely positive maps}

Let us consider a tensor product of two $d$-dimensional Hilbert spaces $\mathcal{H}=\mathcal{H}_1^d\otimes\mathcal{H}_2^d$, and then define the maximally entangled state $\ket{\Phi^{+}}\bra{\Phi^{+}}$ on $\mathcal{H}$ as
\begin{equation*}
\ket{\Phi^{+}}\bra{\Phi^{+}} = \frac{1}{d}\sum_{i,j=1}^{d} \ket{i}\bra{j}\otimes\ket{i}\bra{j}
\end{equation*}
We introduce a completely positive linear map $\Lambda : \mathcal{H}_1^d \rightarrow \mathcal{H}_3^{d'}$, and define the Choi--Jamio\l kowski operator $J(\Lambda) : \mathcal{H}_1^d\otimes\mathcal{H}_2^d \rightarrow \mathcal{H}_3^{d'}\otimes\mathcal{H}_2^d$ as the operator which applies $\Lambda$ to the first half of the maximally entangled state $\ket{\Phi^{+}}\bra{\Phi^{+}}$:
\begin{equation*}
J(\Lambda) = \frac{1}{d}\sum_{i,j=1}^{d} \Lambda(\ket{i}\bra{j})\otimes\ket{i}\bra{j}.
\end{equation*}
Choi's theorem then states that $\Lambda$ is completely positive if and only iff
$J(\Lambda)$ is positive semidefinite. We also have that $\Lambda$ is a trace-preserving map if and only if $\Tr_{\mathcal{H}_3^{d'}}(J(\Lambda)) = \mathbb{1}_{\mathcal{H}_2^d}$ \cite{MVW:tqc12,W:LN11,VB:SIAM96}. These properties are implemented as constraints in the optimization problem from ($\ref{bobby}$).

\subsection*{A.2 : Proof of Equation (\ref{luke})}

For a completely positive trace-preserving linear map $\Lambda:\mathcal{H}_1^d \rightarrow \mathcal{H}_3^{d'}$ and its associated Choi--Jamio\l kowski operator $J(\Lambda)$, and $\ket{\psi_1}\in\mathcal{H}_1^d$, $\ket{\psi_3}\in\mathcal{H}_3^{d'}$, we can write
\begin{align*}
\bra{\psi_3}\otimes&\bra{\overline{\psi_1}} J(\Lambda)  \ket{\psi_3}\otimes\ket{\overline{\psi_1}} \\
&= \bra{\psi_3}\otimes\bra{\overline{\psi_1}}
  \left(\smashoperator[r]{\sum_{i,j=1}^{d} }\Lambda(\ket{i}\bra{j})\otimes\ket{i}\bra{j}\right)
  \ket{\psi_3}\otimes\ket{\overline{\psi_1}}\\
&= \smashoperator{\sum_{i,j=1}^{d}} \bra{\psi_3}\Lambda(\ket{i}\bra{j})\ket{\psi_3}\otimes\braket{\overline{\psi_1}|i}\braket{j|\overline{\psi_1}}\\
&= \smashoperator{\sum_{i,j=1}^{d}} \bra{\psi_3}\Lambda(\ket{i}\bra{j})\ket{\psi_3}\otimes\braket{i|\psi_1}\braket{\psi_1|j}\\
& = \sum_{\mathclap{i,j=1}}^{d} \bra{\psi_3}\Lambda(\psi_{1,ij}\ket{i}\bra{j})\ket{\psi_3}\\
&=\bra{\psi_3}\Lambda(\ket{\psi_1}\bra{\psi_1})\ket{\psi_3},
\end{align*}
where we have defined the scalar $\psi_{1,ij}:=\braket{i|\psi_1}\braket{\psi_1|j}$.

\section*{Appendix B : Non phase-randomized and phase-randomized coherent state expressions.}

\subsection*{B.1 : Non phase-randomized states}

We may write the coherent states $\ket{\alpha_k}= \ket{i^k \frac{\alpha}{\sqrt{2}}}$ from $\mathcal{H}_{\text{mint}}$ in a four-dimensional orthonormal basis $\{\ket{\phi_i}\}$ as
\begin{alignat*}{5}
    \ket{\alpha_0} &=  C_{0}\ket{\phi_0}+&C_{1}\ket{\phi_1}+&C_{2}\ket{\phi_2}+&C_{3}\ket{\phi_3}  \\
    \ket{\alpha_1} &= C_{0}\ket{\phi_0}+i&C_{1}\ket{\phi_1}-&C_{2}\ket{\phi_2}-i&C_{3}\ket{\phi_3}\\
    \ket{\alpha_2} &= C_{0}\ket{\phi_0}-&C_{1}\ket{\phi_1}+&C_{2}\ket{\phi_2}-&C_{3}\ket{\phi_3}\\
    \ket{\alpha_3} &= C_{0}\ket{\phi_0}-i&C_{1}\ket{\phi_1}-&C_{2}\ket{\phi_2}+i&C_{3}\ket{\phi_3}
\end{alignat*}
where
\begin{align*}
    C_0 &= \frac{e^{-\frac{|\alpha|^2}{4}}}{\sqrt{2}}\sqrt{\cosh{\frac{\alpha^2}{2}}+\cos{\frac{\alpha^2}{2}}}\\
    C_1 &= \frac{e^{-\frac{|\alpha|^2}{4}}}{\sqrt{2}}\sqrt{\sinh{\frac{\alpha^2}{2}}+\sin{\frac{\alpha^2}{2}}}\\
    C_2 &= \frac{e^{-\frac{|\alpha|^2}{4}}}{\sqrt{2}}\sqrt{\cosh{\frac{\alpha^2}{2}}-\cos{\frac{\alpha^2}{2}}}\\
    C_3 &= \frac{e^{-\frac{|\alpha|^2}{4}}}{\sqrt{2}}\sqrt{\sinh{\frac{\alpha^2}{2}}-\sin{\frac{\alpha^2}{2}}}
\end{align*}

\subsection*{B.2 : Phase-randomized states}

We may express the four states in our protocol as:
\begin{equation*}
\Ket{e^{i\phi}\frac{\alpha}{\sqrt{2}}}\otimes\Ket{e^{i(\phi+\theta)}\frac{\alpha}{\sqrt{2}}},
\label{globalphase}
\end{equation*}
with global phase $\phi=0$ and relative phase $\theta\in\{0,\frac{\pi}{2},2\pi,\frac{3\pi}{2}\}$. This implies that an adversary must access $\theta$ to unveil the information encoded in the states. Phase randomization scrambles the global phase reference by allowing $\phi$ to take values from $[0,2\pi]$ uniformly at random instead of a single value. By considering the state $\ket{e^{i\phi}\alpha}$ and integrating over all possible values of $\phi$, the adversary sees a classical mixture of Fock states given by \cite{LP:calt05}:
\begin{equation*}
    \frac{1}{2\pi}\int_{0}^{2\pi} \ket{\sqrt{\mu} e^{i\phi}}\bra{\sqrt{\mu} e^{i\phi}}d\phi = e^{-\mu}\sum_{n=0}^{\infty}\frac{\mu^n}{n!}\ket{n}\bra{n},
\end{equation*}
where $\mu=|\alpha|^2$ is the average photon number, and $\ket{n}$ are the photon number states. As the coherent superpositions of number states vanish, the security proof may simply proceed according to the result of quantum non demolition (QND) photon number measurements. If there is no photon in the state, then there is no information. If there is $1$ photon, then the qubit security proof may be applied.
If there are more than $2$ photons in the pulse, perfect cheating is possible,
since one photon can be sent to a terminal 1 and another to terminal 2.
For our protocol, this allows us to express the phase randomized states $\rho_k$ in a $7$-dimensional orthonormal basis $\{\ket{v},\ket{q_0},\ket{q_1},\ket{m_0},\ket{m_1},\ket{m_2},\ket{m_3}\}$, where $\ket{v}$ is the vacuum state, $\ket{q_0}$ and $\ket{q_1}$ span a qubit space, and $\ket{m_i}$ constitute the four orthogonal outcomes which materialize the four perfectly distinguishable states in the multiphoton subspace. Our four phase-randomized coherent states may then be written as the following density matrices :
\begin{alignat*}{4}
    \rho_0 &= p_0(\mu)\ketbra{v}{v}+&p_1(\mu)\ketbra{+}{+}+p_m(\mu)&\ketbra{m_0}{m_0}\\
    \rho_1 &= p_0(\mu)\ketbra{v}{v}+&p_1(\mu)\ketbra{+i}{+i}+p_m(\mu)&\ketbra{m_1}{m_1}\\
    \rho_2 &= p_0(\mu)\ketbra{v}{v}+&p_1(\mu)\ketbra{-}{-}+p_m(\mu)&\ketbra{m_2}{m_2}\\
    \rho_3 &= p_0(\mu)\ketbra{v}{v}+&p_1(\mu)\ketbra{-i}{-i}+p_m(\mu)&\ketbra{m_3}{m_3},
\end{alignat*}
where $\ket{+}$,$\ket{+i}$, $\ket{-}$, $\ket{-i}$ are the usual $\sigma_x$ and $\sigma_y$ eigenstates in the qubit space spanned by $\ket{q_i}$ and the Poisson distribution coefficients are given by
\begin{align*}
    p_0(\mu)&= e^{-\mu},&
    p_1(\mu) &= \mu e^{-\mu},&
    p_m(\mu) &= 1-(1+\mu)e^{-\mu}.
\end{align*}

\section*{Appendix C : Extension of SDP ($\ref{bobby}$) to $n$ parallel repetitions.}

Semidefinite programming presents a dual structure, which associates a dual maximization problem to each primal minimization problem \cite{,W:LN11,VB:SIAM96}. The optimal value of the primal problem then upper bounds the optimal value of the dual problem, and the optimal value of the dual problem lower bounds that of the primal problem. This property is known as $\textit{weak duality}$. We also note that a single problem may admit several \textit{feasible solutions}, \emph{i.e.}, operators which satisfy all the constraints. The \textit{optimal solution} is the feasible solution which optimizes the objective function (the quantity we aim to minimize or maximize). In our setting, SDP (\ref{bobby}) may be labelled as the primal problem. The aim of this section is to extend SDP (\ref{bobby}) to a credit card containing $n$ states, and to derive its corresponding dual problem. In order to show that the adversary does not gain any advantage in correlating the $n$ states to better succeed, we will first show that a tensor product of $n$ optimal solutions of (\ref{bobby}) is a feasible solution to this new primal SDP.
We then have to show that there also exists a feasible solution for the associated dual problem which yields the same optimal value as that of the primal.
Such a feature is known as $\textit{strong duality}$, and implies that these feasible solutions are both the $\textit{optimal solutions}$ to the primal and dual problem, respectively.
We failed to prove this strong duality analytically, but as shown below, numerical evidence indicate it should hold.

To generalize the loss and error operators to the $n$ parallel repetition case, we introduce the projector $\mathcal{P}(n,j,\mathcal{C})$ which, given a collection $\mathcal{C}$ of $n$ quantum states living in Hilbert space $\mathcal{H}^{(n)}$, projects onto $j\leqslant n$ elements of $\mathcal C$ and the orthogonal subspace of the $(n-j)$ other elements. More formally, this operator is defined as
\begin{equation*}
    \mathcal{P}(n,j,\mathcal{C}) = \sum_{s(j)} \bigotimes_{i=0}^{n-1} \left[s_i(j) \mathcal{C}_i + \overline{s_i(j)} (\mathbb{1}-\mathcal{C}_i)\right],
\end{equation*}
where $\mathcal{C}_i$ is the $i$-th quantum state of $\mathcal{C}$ and $s_i(j)$ is the $i$-th element of a binary string $s(j)$ of length $n$ which contains $(n-j)$ zeros. The summation then runs over all $\binom{n}{j}$ possible $s(j)$     strings. Considering a new adversarial cloning map $\Lambda^{(n)}$ from the original $n$-state credit card living in $\mathcal{H}^{(n)}_{\text{mint}}$ to a duplicated credit card space $\mathcal{H}^{(n)}_1\otimes\mathcal{H}^{(n)}_2$, the new loss operators may then be written as:
\begin{gather*}
\begin{multlined}
L^{(n)}_1(\mu) = \frac{1}{4^n}\sum_{j=1}^{n}\smashoperator[r]{\sum_{k_1\cdots k_n=0}^{3}}\tfrac{j}{n} \mathcal{P}\left(n,j,\mathcal{C}^{(\varnothing,n)}\right) \otimes\mathbb{1}_{\mathcal{H}^{(n)}_2}\\
  \otimes\left(\ket{\overline{\alpha_{k_1}}}\bra{\overline{\alpha_{k_1}}}\otimes\cdots \otimes \ket{\overline{\alpha_{k_n}}}\bra{\overline{\alpha_{k_n}}} \right) \\
\end{multlined}\\
\begin{multlined}
L^{(n)}_2(\mu) = \frac{1}{4^n} \sum_{j=1}^{n}\smashoperator[r]{\sum_{k_1\cdots k_n=0}^{3}}
  \mathbb{1}_{\mathcal{H}^{(n)}_1}\otimes\tfrac{j}{n}\mathcal{P}\left(n,j,\mathcal{C}^{(\varnothing,n)}\right)\\
  \otimes\left(\ket{\overline{\alpha_{k_1}}}\bra{\overline{\alpha_{k_1}}}\otimes\cdots \otimes \ket{\overline{\alpha_{k_n}}}\bra{\overline{\alpha_{k_n}}} \right),
\end{multlined}
\end{gather*}
where $\mathcal{C}^{(\varnothing,n)}=\{\ket{\varnothing}\bra{\varnothing}\}^n$. The factors $\frac{j}{n}$ ensure that the total sum is normalized, as we are dealing with probabilities and not events. The new error operators read:
\begin{gather*}
\begin{multlined}
  E^{(n)}_1(\mu) = \frac{1}{4^n}\sum_{j=1}^{n}\smashoperator[r]{\sum_{k_1\cdots k_n=0}^{3}} \tfrac{j}{n}\mathcal{P}\left(n,j,\mathcal{C}^{(k_1,.,k_n)}\right)  \otimes\mathbb{1}_{\mathcal{H}^{(n)}_2}\\
    \otimes\left(\ket{\overline{\alpha_{k_1}}}\bra{\overline{\alpha_{k_1}}}\otimes\cdots \otimes \ket{\overline{\alpha_{k_n}}}\bra{\overline{\alpha_{k_n}}} \right)
\end{multlined}\\
\begin{multlined}
  E^{(n)}_2(\mu) =
    \frac{1}{4^n} \sum_{j=1}^{n}\smashoperator[r]{\sum_{k_1\cdots k_n=0}^{3}} \mathbb{1}_{\mathcal{H}^{(n)}_1}\otimes\tfrac{j}{n}\mathcal{P}\left(n,j,\mathcal{C}^{(k_1,.,k_n)}\right)\\
    \otimes\left(\ket{\overline{\alpha_{k_1}}}\bra{\overline{\alpha_{k_1}}}\otimes\cdots \otimes \ket{\overline{\alpha_{k_n}}}\bra{\overline{\alpha_{k_n}}} \right),
\end{multlined}
\end{gather*}
where $\mathcal{C}^{(k_1,\dots,k_n)}=\{\frac{1}{2}\ket{\beta^{\perp}_{k_1}}\bra{\beta^{\perp}_{k_1}},\dots,\frac{1}{2}\ket{\beta^{\perp}_{k_n}}\bra{\beta^{\perp}_{k_n}}\}$. For a credit card containing $n$ states, problem (\ref{bobby}) may then be recast as:
\begin{equation}
\begin{aligned}
\min & \Tr\left(L^{(n)}_1(\mu) J(\Lambda^{(n)}) \right)\\
\text{s.t.}  & \Tr_{\mathcal{H}^{(n)}_1\otimes\mathcal{H}^{(n)}_2}\left(J(\Lambda^{(n)})\right) = \mathbb{1}_{\mathcal{H}^{(n)}_{\text{mint}}} \\
&\Tr\left(E^{(n)}_1(\mu) J(\Lambda^{(n)}) \right) = e \\
& \Tr\left(E^{(n)}_1(\mu) J(\Lambda^{(n)}) \right) \geqslant \Tr\left(E^{(n)}_2(\mu) J(\Lambda^{(n)}) \right) \\
& \Tr\left(L^{(n)}_1(\mu) J(\Lambda^{(n)}) \right) \geqslant\Tr\left(L^{(n)}_2(\mu) J(\Lambda^{(n)}) \right) \\
& J(\Lambda^{(n)}) \geqslant0
\end{aligned}\label{bobby_n}
\end{equation}
To derive the dual problem associated with (\ref{bobby_n}), we first note that we can replace all inequalities by equalities (except the last semidefinite positive constraint) without loss of generality. This is due to the fact that the adversary can always symmetrize the probabilities by increasing the error rate or losses on card 2 to make them equal to those on card 1. The right hand side elements of the constraints from ($\ref{bobby_n}$) may then be gathered in a $(4^{2n}+3)$-dimensional column vector $\vec{b}^{(n)}$. The first three elements read $(e,e,0)$, and correspond to the value of $\Tr\left(E^{(n)}_1(\mu) J(\Lambda^{(n)}) \right)$, $\Tr\left(E^{(n)}_2(\mu) J(\Lambda^{(n)}) \right)$ and $\Tr\left(\left(L^{(n)}_1(\mu)-L^{(n)}_2(\mu)\right)J(\Lambda^{(n)}) \right)$, respectively. The $4^{2n}$ other elements, corresponding to the first, trace-preserving constraint of (\ref{bobby_n}), may be written as the vector representation of the identity over space $\mathcal{H}^{(n)}_{\text{mint}}$. The vector representation $vec(O)$ of an operator $O$ is obtained through the following isomorphism \cite{DCP:CUP17} :
\begin{equation}
\sum_{ij=1}^{d} O_{ij}\ket{i}\bra{j} \rightarrow \sum_{ij=1}^{d} O_{ij}\ket{i}\otimes\ket{j}
\end{equation}
%

\begin{table}
\begin{ruledtabular}
\begin{tabular}{ccccccc}
 $e$& $10^{-6}$ & $10^{-3}$ & 0.01 & 0.02 & 0.05 & 0.10  \\
 $\mu$ & 0.01 & 0.05 & 0.10 & 0.50 & 1.00 & 2.00
\end{tabular}
\end{ruledtabular}
\caption{\label{tab:dualparam} Numerical solutions of the
   dual problem \eqref{dual} were found for all possible combinations of the above $e$ and $\mu$ values, using the \textit{SDPT3} solver from the \textit{CVX} software with its default numerical precision ($10^{-9}$). These solutions obeyed the constraints \eqref{conditions} within numerical accuracy and all present a duality gap of order $10^{-9}$. We note that, when $e<10^{-6}$, the solver struggles to find an accurate solution for some low values of $\mu$, as it fails to decrease the duality gap to less than $10^{-7}$. The inaccurate optimal dual solutions are nevertheless close to the accurate primal optimal solutions within $10^{-4}$. }

\end{table}

 The dual problem then maximizes the overlap of variable $\vec{d}^{(n)}$ with constraint vector $\vec{b}^{(n)}$ as:
\begin{equation}
\begin{aligned}
\max\vec{b}^{(n) T} &\vec{d}^{(n)}\\
\text{s.t. } d^{(n)}_1 &E^{(n)}_1(\mu)  +d^{(n)}_2 E^{(n)}_2(\mu)\\
  &+d^{(n)}_3 \left(L^{(n)}_1(\mu)-L^{(n)}_2(\mu)\right) \\ &+\mathbb{1}_{\mathcal{H}^{(n)}_1\otimes\mathcal{H}^{(n)}_2}\otimes D^{(n)} - L^{(n)}_1(\mu) \leqslant 0,
\label{dual}
\end{aligned}
\end{equation}
where $D^{(n)}$ is a $4^{n}\times4^n$ matrix containing the elements $d^{(n)}_4$ to $d^{(n)}_{4^{2n}+3}$ arranged in order left to right, top to bottom. The objective function reads $\vec{b}^{(n) T} \vec{d}^{(n)} = ed_1^{(n)}+ed_2^{(n)}+\Tr(D^{(n)})$. We note that a tensor product of optimal solutions $J(\Lambda^{(n)}) = \bigotimes_{j=1}^{n} J(\Lambda)$ represents a feasible solution to primal problem (\ref{bobby_n}), as it satisfies all the constraints. We label the associated primal objective function value as $s_p^{(n)}$, and remark that $s_p^{(n)}=s_p^{(1)}=f_{d}$ for all $n$. We then search for a feasible solution $\vec{d}^{(n)}$ to the dual problem (\ref{dual}) which allows to achieve $s_p^{(n)}=s_d^{(n)}$, where $s_d^{(n)}$ is the dual objective function value.
While we were not able to find a generic analytical solution to this problem, we have always found a numerical solution $\vec{d}^{(n)}$ for a representative set of parameters $\mu$ and $e$ (specified in Table \ref{tab:dualparam}), satisfying
\begin{equation}
\begin{aligned}
        d^{(n)}_1&, d^{(n)}_2 < 0     &D^{(n)}_{ij} &= 0 \text{ for } i\neq j\\
        d^{(n)}_3 &= 0.5       &\Tr(D^{(n)})  &= s_p^{(n)}-(d^{(n)}_1+d^{(n)}_2)e,
\label{conditions}
\end{aligned}
\end{equation}
and presenting a duality gap of order $10^{-9}$. Furthermore, adding the last condition as constraint to the SDP does not change the optimal value (within $10^{-4}$ error, due to the fact the the value of $s_p^{(n)}$ is a numerical primal optimal value which is rounded up when added as a constraint in the dual problem).
The conditions on $d^{(n)}_1,d^{(n)}_2,d^{(n)}_3$ enforce the following expression of the dual constraint:
\begin{multline}
-\abs{d^{(n)}_1}E^{(n)}_1(\mu)-\abs{d^{(n)}_2} E^{(n)}_2(\mu)-\tfrac{1}{2}L^{(n)}_1(\mu)\\
-\tfrac{1}{2}L^{(n)}_2(\mu) + \mathbb{1}_{\mathcal{H}^{(n)}_1\otimes\mathcal{H}^{(n)}_2}\otimes D^{(n)} \leqslant 0
\label{constraint}
\end{multline}
 Since the error and loss operators are all positive semidefinite, then it follows that the sum of the first four terms in (\ref{constraint}) is a negative semidefinite operator. Numerically, it appears to be possible to satisfy (\ref{constraint}) by choosing appropriately the diagonal elements of $D^{(n)}$,
 and we conjecture it is always possible.

In conclusion, we have found two feasible solutions such that $s_p^{(n)}=s_d^{(n)}$, and strong duality holds for problems (\ref{bobby_n}) and (\ref{dual}),
at least up to numerical precision. The optimal solution to the primal problem for $n$ states can therefore be written as a tensor product of optimal solutions to the primal problem for $n=1$ state. This implies that the adversary does not gain any advantage in correlating the states in the card when performing an attack against a trusted terminal without phase randomization. A similar approach works to prove strong duality for the untrusted terminal case, and we conjecture that this method also works for both scenarios with phase-randomized states.

\section*{Appendix D : Optimal adversarial strategy with qubits in an untrusted terminal scenario.}

A simple strategy corresponding to the error rate $e=1/8$ for a state encoded in the basis $b$ is :
\begin{description}
\item[$c_i = c_j$] Adopt the honest strategy and duplicate the classical outcome.\\
Success probability: $1$
\item[$c_i \neq c_j$] Pick a basis $b$ (or $\overline{b}$) at random, measure the state in this
basis, and send the classical outcome to answer challenge $c_b$ (or $c_{\overline{b}}$). Send a random measurement outcome to the other challenge $c_{\overline{b}}$ (or $c_b$). If the correct basis $b$ was picked, then the adversary succeeds with probability $1$. If the wrong basis $\overline{b}$ was picked, then the success probability is $\frac{1}{2}$. \\Success probability : $\frac{1}{2}\times 1 +\frac{1}{2}\times\frac{1}{2}$ = $\frac{3}{4}$.
\end{description}
Since the bank will ask each of these challenge combinations with probability $\frac{1}{2}$, then we have a total success probability $\frac{7}{8}$, which yields $e=\frac{1}{8}$.

\section*{Appendix E : Optimal error rate}

If one wishes to minimize the error rate given a fixed honest loss rate, then one may cast the following SDP, which has a similar structure to (\ref{bobby}) :

\begin{align*}
\min\quad& \Tr\left(E_1(\mu) J(\Lambda) \right) \displaybreak[1]\\
\text{s.t. }&  \Tr_{\mathcal{H}_1\otimes\mathcal{H}_2}\left(J(\Lambda)\right) = \mathbb{1}_{\mathcal{H}_{\text{mint}}} \\
    &\Tr\left(E_1(\mu) J(\Lambda) \right) \geqslant \Tr\left(E_2(\mu) J(\Lambda) \right)
     \addtocounter{equation}{1}\tag{\theequation} \label{jacky} \displaybreak[1]\\
    &\Tr\left(L_1(\mu) J(\Lambda) \right) \leqslant e^{-\eta_d\mu} \displaybreak[1]\\
    &\Tr\left(L_2(\mu) J(\Lambda) \right) \leqslant e^{-\eta_d\mu} \\
    &J(\Lambda) \geqslant 0
\end{align*}

The first constraint imposes that $\Lambda$ is trace-preserving, the second imposes that the error on card $1$ is greater or equal to that on card $2$, the third and fourth impose that the losses on each card are smaller or equal to the honest expected losses $f_h$, and the fifth imposes that $\Lambda$ is completely positive. Tables \ref{tab:a123} give optimal numerical solutions to (\ref{jacky}) for different scenarios, varying the phase-randomization, detection efficiencies, terminal trust,
losses and average photon number $\mu$.

\begin{table}[H]
\begin{ruledtabular}
\begin{tabular}{cccc}
 $\mu$ & $e$, trusted & $e$, untrusted & $f_h$  \\
\hline
\multicolumn{4}{l}{Non phase-randomized, $\eta_{d}=100\%$}\\
0.05 & 0.1\% & 0.1\% & 95.1\% \\
0.10 & 0.3\% & 0.1\% & 90.5\% \\
0.15 & 0.4\% & 0.1\% & 86.1\% \\
0.25 & 0.5\% & 0 \% & 77.9\% \\
0.55 & 0.7\% & 0 \% & 57.7\% \\
\hline
\multicolumn{4}{l}{Phase-randomized, $\eta_{d}=100\%$}\\
0.50 & 2.2\% & 1.1\% & 60.1\% \\
0.75 & 2.6\% & 1.3\% & 47.2\% \\
1.00 & 2.7\% & 1.3\% & 36.8\% \\
1.25 & 2.6\% & 1.3\% & 28.7\% \\
1.50 & 2.4\% & 1.2\% & 22.3\% \\
\hline
\multicolumn{4}{l}{Phase-randomized, $\eta_{d}=80\%$}\\
0.40 & 1.1\% & 0.2\% & 72.6\% \\
0.60 & 1.4\% & 0.2\% & 61.9\% \\
0.80 & 1.5\% & 0.2\% & 52.7\% \\
1.00 & 1.5\% & 0.1\% & 44.9\% \\
1.20 & 1.5\% & 0.1\% & 38.3\% \\
\end{tabular}
\end{ruledtabular}
\caption{\label{tab:a123} Optimal numerical solutions to \eqref{jacky}\ in various scenarios.}
\end{table}

\end{document}